\documentclass{emulateapj}

\shorttitle{Chandra Reveals X-ray Jets in 3C~353}
\shortauthors{Kataoka et al.}
\usepackage{times}
\begin{document}

\title{{\it Chandra} Reveals Twin X-ray Jets in the Powerful FR~II Radio
Galaxy 3C~353}

\author{J. Kataoka\altaffilmark{1}, 
\L. Stawarz\altaffilmark{2,\,3}, 
D.E. Harris\altaffilmark{4}, 
A. Siemiginowska\altaffilmark{4}, 
M. Ostrowski\altaffilmark{3}, \\
M.R. Swain\altaffilmark{5}, 
M.J. Hardcastle, J.L. Goodger\altaffilmark{6}, 
K. Iwasawa\altaffilmark{7}, 
and P.G. Edwards\altaffilmark{8}}

\altaffiltext{1}{Department of Physics, Tokyo Institute of Technology, 2-12-1 Ohokayama, Meguro, Tokyo, 152-8551, Japan}
\email{kataoka@hp.phys.titech.ac.jp}
\altaffiltext{2}{Kavli Institute for Particle Astrophysics and Cosmology, Stanford University, Stanford, CA 94305, USA}
\altaffiltext{3}{Obserwatorium Astronomiczne, Uniwersytet
Jagiello\'nski, ul. Orla 171, 30-244 Krak\'ow, Poland}
\altaffiltext{4}{Harvard-Smithsonian Center for Astrophysics, 60 Garden St., Cambridge, MA 02138}
\altaffiltext{5}{Jet Propulsion Laboratory, California Institute of Technology, 4800 Oak Grove Drive, Pasadena, CA 91109}
\altaffiltext{6}{University of Hertfordshire, College Lane, Hatfield, Hertfordshire AL 10 9AB, UK}
\altaffiltext{7}{INAF-Osservatorio Astronomico di Bologna, Via Ranzani,
1, 40127, Bologna, Italy}
\altaffiltext{8}{Australia Telescope National Facility, CSIRO, Locked Bag 194, Narrabri NSW 2390, Australia}

\begin{abstract}
We report X-ray imaging of the powerful FR~II radio galaxy 3C~353
 using the {\it Chandra} X-ray Observatory. 3C~353's two
 $4$\arcsec-wide and $2$\arcmin-long jets allow us to study in detail
 the internal structure of the large-scale relativistic outflows at
 both radio and X-ray photon energies with the sub-arcsecond spatial
 resolution provided by the VLA and {\it Chandra} instruments.
 In a $90$\,ks {\it Chandra} observation, we have detected X-ray
 emission from most radio structures in 3C~353, including the nucleus,
 the jet and the counterjet, the terminal jet regions (hotspots), and
 one radio lobe. We show that the detection of the X-ray emission
 associated with the radio knots and counterknots, which is most
 likely non-thermal in origin, puts several crucial constraints on the
 X-ray emission mechanisms in powerful large-scale jets of quasars and
 FR~II sources. In particular, we show that this detection is
 inconsistent with the inverse-Compton model proposed in the
 literature, and instead implies a synchrotron origin of the X-ray jet
 photons. We also find that the width of the X-ray counterjet is
 possibly narrower than that measured in radio bands, that the
 radio-to-X-ray flux ratio decreases systematically downstream along
 the jets, and that there are substantial (kpc-scale) offsets between
 the positions of the X-ray and radio intensity maxima within each
 knot, whose magnitudes increase away from the nucleus. We discuss all
 these findings in the wider context of the physics of extragalactic
 jets, proposing some particular though not definitive solutions or
 interpretations for each problem. In general, we find that the
 synchrotron X-ray emission of extragalactic large-scale jets is not
 only shaped by the global hydrodynamical configuration of the
 outflows, but is also likely to be very sensitive to the microscopic
 parameters of the jet plasma. A complete, self-consistent model for
 the X-ray emission of extragalactic jets still remains elusive.

\end{abstract}


\section{Introduction}
\label{sec:intro}

Extragalactic jets constitute the longest collimated structures in the
Universe. They transport huge amounts of energy from the nuclei of active
galaxies out to kpc or Mpc distances, significantly affecting the properties
of the surrounding intracluster/intergalactic medium. Jets have been
extensively studied in the radio domain on different scales since the very
beginning of the development of modern radio interferometers (Bridle \&
Perley 1984, Begelman et al. 1984). More recently, the excellent spatial
resolution of the {\it Chandra} X-ray Observatory (and, to a lesser extent,
of other X-ray satellites like {\it XMM-Newton}) has allowed us to
image large-scale structures in powerful extragalactic radio sources
at X-ray frequencies as well, and thus has opened a new era in studying the
high energy emission of these objects. More than 100 radio-loud AGNs are
now known to possess X-ray counterparts to their radio jets, hotspots or
lobes on kpc-to-Mpc scales (e.g., Harris \& Krawczynski 2002; 2006,
Stawarz 2003, Hardcastle et al. 2004, Sambruna et al. 2004, Kataoka \&
Stawarz 2005, Marshall et al. 2005a, Croston et al. 2005, Tavecchio et al. 
2005, Hardcastle 2006, and references therein). The X-ray emission observed 
from the extended lobes is well understood and modeled in terms of the
inverse Comptonization of the cosmic microwave background photons by the
low-energy electrons (IC/CMB; see the discussion in Kataoka \& Stawarz
2005, and Croston et al. 2005), providing strong evidence for
approximate energy equipartition between the radiating electrons and the
lobe magnetic field (with the particle pressure dominating over the
magnetic pressure by up to one order of magnitude). However, the origin of the
$0.1-10$\,keV radiation detected from the large-scale jets and hotspots
is still widely debated and, to some extent, controversial.

The X-ray emission of the terminal regions (the hotspots) of
\emph{powerful} jets is consistent with the synchrotron self-Compton
(SSC) emission model, in which radio-emitting electrons accelerated at
the terminal shock inverse-Compton-scatter synchrotron radio photons
to keV energies (e.g., Hardcastle et al. 2004, Kataoka \& Stawarz
2005, and references therein). Modeling of X-rays from this emission
process allows us to extract several crucial jet parameters. In
particular, the observations imply rough equipartition of the energy
densities stored in radiating electrons, $U_{\rm e}$, and magnetic
field, $U_{\rm B} \lesssim U_{\rm e}$, and it has been argued that
they indicate a dynamical role for non-relativistic protons in the
outflow (see Stawarz et al. 2007 for the case of the radio galaxy
Cygnus A). On the other hand, the X-ray emission of the hotspots in
\emph{low-power} jets is in disagreement with the predictions of the
simple SSC model (see the discussion in Hardcastle et al. 2004). The
complex X-ray morphology and spectral character of the broad-band
hotspot emission in objects like 3C~227 or 3C~327 suggest instead that
we are seeing synchrotron emission from high-energy electrons
accelerated continuously in the extended and turbulent jet termination
regions (Hardcastle et al. 2007a, Fan et al. 2008). The origin of the
bright X-ray knots within the jets themselves (hereafter `jet knots')
is also a matter of a debate. In nearby, low-power FR~I sources (like
M~87 and Centaurus~A), the typical radio-to-X-ray spectra of the jet
knots are consistent with a single smoothly broken power-law
continuum, indicating synchrotron radiation from a single electron
population extending up to the highest energies ($\gamma \equiv E_{\rm
e} / m_{\rm e} c^2 \sim 10^8$; e.g., Marshall et al. 2002, Hardcastle
et al. 2003). Yet the particular acceleration processes involved,
the multi-component character of the high-energy emission, and the
main factors determining the spectral shape of the broad-band jet
emission, are far from being understood (see the discussions in, e.g.,
Kataoka et al. 2006, Laing et al. 2006a, Honda \& Honda 2007).

The most controversial issue, however, is the origin of the intense
X-ray emission detected from knots in powerful quasar jets, such as
PKS~0637$-$752 or 3C~273 (Schwartz et al. 2000, Marshall et al. 2001,
respectively). Here the X-ray knot spectra are much brighter than
expected from a simple extrapolation of the radio-to-optical
synchrotron continua, indicating that an additional or separate
spectral component dominates the jet's radiative output at high (X-ray)
photon energies. Very often, this emission is modeled in terms of
inverse-Comptonization of the CMB photon field by low-energy ($\gamma
\lesssim 10^3$) electrons (Tavecchio et al. 2000, Celotti et al.
2001), which typically requires highly relativistic jet bulk
velocities on kpc-Mpc scales in order to maintain rough equipartition
between electrons and magnetic field and to fulfill the minimum power
condition. The derived jet bulk Lorentz factors $\Gamma_{\rm jet}
\gtrsim 10$ are comparable to those inferred for the pc-scale jets;
(Harris \& Krawczynski 2002, Sambruna et al. 2004, Marshall et al.
2005a). Unfortunately, the global dynamics of large-scale jets in
powerful sources are largely unknown, and different arguments in favor
of and against highly relativistic bulk velocities on kpc-Mpc scales
(obtained by means of the analysis of the radio or optical jet
emission; e.g., Wardle \& Aaron 1997, Hardcastle et al. 1999, Scarpa
\& Urry 2002), are not conclusive. Thus, the IC/CMB model may be
considered to be strong support for the idea that these outflows do
indeed propagate from sub-pc scales with little energy dissipation,
efficiently transporting energy in the form of bulk kinetic motion,
and depositing it far away from the active nuclei (Tavecchio et al.
2004; 2007, Sambruna et al. 2006; but see the discussion in Hardcastle
2006). This important conclusion is often claimed to be consistent
with the observed one-sideness of the extragalactic X-ray jets
detected by {\it Chandra}, since jet one-sideness is a simple and
natural consequence of relativistic beaming.


On the other hand, if there is significant beaming in powerful jets on
large scales, then the detection of bright X-ray jet emission from
FR~II radio galaxies, which are believed to be analogous systems to
radio loud quasars but to be viewed with the jets at large angles to
the line of sight, should be considered as unlikely. Such emission
has, however, been detected in several objects (e.g., 3C~303, 3C~15,
Pictor~A, or 3C~403; see Kataoka et al. 2003a,b, Hardcastle \& Croston
2005, and Kraft et al. 2005, respectively). Obviously, detection of
any X-ray counterjet would be of primary importance in this respect,
since it would automatically exclude significant beaming, and thus
impose very severe constraints on the jet emission models. In FR~I
sources, where the emission mechanism is generally supposed to be
synchrotron, possible or likely detections of X-ray counterjets have
been reported in 3C~270 and Centaurus~A (Chiaberge et al. 2003,
Hardcastle et al. 2007b, respectively). In the latter case the
counterjet detection is almost certain, since \emph{extended} X-ray
emission coincident with the \emph{extended} radio features in the
receding jet rules out a claim that all of the observed X-ray
counterparts to the radio counter-knots may be due to chance
coincidence of the jet-related radio features with X-ray binaries of
the Centaurus~A host galaxy. However, in more powerful sources, there
has been no definitive counterjet detection to date. Possible
counterjets have been reported in the intermediate FR~I/FR~II object
4C~29.30 (Sambruna et al. 2004), as well as in the broad-line FR~II
radio galaxy Pictor~A (Hardcastle \& Croston 2005). In these cases, it
is unclear whether the X-ray emission is non-thermal in nature,
because with the available very limited photon statistics any detailed
spectral analysis is impossible, and in addition it is rather difficult
to claim one-to-one morphological correspondence between the X-ray and
radio structures (especially since the counterjet in the prototype
FR~II source Pictor~A is not detected at radio frequencies). In X-ray
maps of the classical double (narrow-line FR~II radio galaxy)
Cygnus~A, linear features aligned --- but not co-spatial --- with the
radio jet and the counterjet have been noted, but their direct
connection with the radio jet plasma is extremely vague (Smith et al.
2002, Steenbrugge et al. 2008). It is then possible that in this, and
some other possible X-ray counterjet sources, the aligned X-ray
structures on the counterjet side are due to thermal radiation of
shocked galactic/intergalactic gas interacting with the jet plasma, as
observed in the distant ($z=2.2$) radio galaxy PKS~1138-262 (Carilli
et al. 2002). We note in this context that Cygnus~A is located in the
very center of a rich cluster environment, while the restarting
4C~29.30 radio galaxy is known to have clear signatures of the
interaction between the radio jets and the ambient line-emitting gas
(van Breugel et al. 1986, Jamrozy et al. 2007). A clear spatial
association between the radio and X-ray jets and the capability to
distinguish non-thermal and thermal emission are the key requirements
for a convincing counterjet detection in a powerful source.

An alternative to the inverse-Compton model as an interpretation of
the X-ray emission of powerful jets involves synchrotron emission from
high-energy electrons, which, in order to be consistent with the
constraints on optical emission, must often either be characterized by
a `non-standard' (concave) energy distribution, or must constitute a
separate population to the ones emitting the radio-to-optical
continuum. It has been noted that such an electron population can
arise due to the continuous and efficient stochastic acceleration
processes expected to take place within the extended jet volumes
(Stawarz et al. 2004), and will be seen preferentially within
turbulent jet boundary layers with significant velocity shear (Stawarz
\& Ostrowski 2002). A very strong support for the synchrotron
hypothesis was recently provided by detailed multiwavelength
observations of the jet in the quasar 3C~273 (Jester et al. 2006,
2007, Uchiyama et al. 2006). The observations showed in particular
that the X-ray spectra are significantly softer than the radio spectra
in most regions of the 3C~273 outflow, and that they are compatible
with extrapolating the X-ray power law down to the polarized (and
therefore synchrotron in origin) UV/optical continuum. These findings
are in strong disagreement with the predictions of the IC/CMB model.
In addition, a detailed analysis of the broad-band emission of several
other quasar jets supports the synchrotron hypothesis, pointing out an important role of the jet velocity structure (consisting
of a fast spine and a slower boundary layer/outer sheath) in shaping
the jet high-energy radiation (e.g., Hardcastle 2006, Jester et al.
2006, Siemiginowska et al. 2007). Such jet velocity structure is in
fact always seen in numerical modeling of the evolution and
propagation of extragalactic relativistic jets (e.g., Aloy et al.
1999, Leismann et al. 2005, Mizuno et al. 2007), since it results
inevitably from the non-linear growth of Kelvin-Helmholtz
instabilities on the jet surface (see the recent studies by Perucho et
al. 2007, Hardee 2007, Meliani \& Keppens 2007, and references
therein). It is not yet established, however, if the jet boundary layers
are indeed the places of the enhanced acceleration of high-energy
particles. Nor is the exact velocity profile at the jet boundaries
precisely known; in fact, there are reasons to believe it may be far
from the widely expected monotonic one, and may even exhibit a sharp
increase in the bulk velocity of the outflow at the jet edges (Aloy \&
Rezzolla 2006, Mizuno et al. 2008). The radiative signatures of such
`anomalous' shear layers, in the context of the high-energy emission
from extragalactic large-scale jets, were recently investigated by
Aloy \& Mimica (2008).

Observational studies of the velocity structure of extragalactic jets
and its relationship to their broad-band emission are hampered by the
difficulties in resolving the outflows transversely, especially at
high energies. Detailed radio studies and sophisticated modeling of
the polarization and total intensity in the radio of several nearby
FR~I sources (like 3C~31, NGC~315, or 3C~296; Laing \& Bridle 2004,
Laing et al. 2006a,b, respectively) do reveal jet radial velocity
structures, and indicate that the flatter-spectrum radio regions are
indeed often associated with the jet boundary shear layers.
Unfortunately, only a few kpc-scale low-power jets, e.g., the one
hosted by the radio galaxy Centaurus~A, can be resolved at X-ray
frequencies by the {\it Chandra} instrument. The Cen A observations show
interestingly limb-brightened X-ray morphology, but
there is no obvious relation with the spectral shape of the
synchrotron X-ray continuum (Hardcastle et al. 2003, 2007b, Kataoka et
al. 2006, Worrall et al. 2008). In particular, the diffuse component
of the Centaurus~A X-ray jet is characterized by a constant spectral
index across the jet (Kataoka et al. 2006), while the X-ray spectra of
isolated knots seem to be steeper within the jet sheath when compared
to the X-ray spectra of the knots within the spine (Worrall et al.
2008). Recently, two `intermediate' FR~I/FR~II jets in the BL Lacertae
objects 3C~371 and PKS~2201+44 were also resolved transversely in
X-rays (Sambruna et al. 2007). The observations indicated a
synchrotron origin of the detected keV photons from all the jet
components, and suggested slightly steeper X-ray spectra of the jet
edges when compared with the jet spines. Interestingly, the width of
the X-ray jets in these two sources is not smaller than the widths of
their optical and radio counterparts. Needless to say, no quasar or
powerful FR~II jet has been resolved till now at X-ray frequencies.
This is true even for the particularly bright and wide 3C~273 jet,
which, although imaged in detail at optical and radio frequencies
(Jester et al. 2005), cannot be resolved transversely at higher photon
energies. For this source, however, the limits to the width of the
X-ray jet are smaller than the sized measured in the optical and radio
bands at the position of the bright knots (Jester et al. 2006), but
not necessarily in the interknot regions (see Marshall et al. 2005b).

The FR II radio galaxy 3C~353 provides one of the best-known examples
of FR II jets that can be resolved at both radio and X-ray
wavelengths. 3C~353 ($z = 0.0304$) is the fourth strongest radio
source in the 3C catalog, with a total flux density $S_{\nu} \approx
57$\,Jy at $1.4$\,GHz, and a projected size $\sim 4.5\arcmin$. It
exhibits hotspots and a pair of large-scale FR~II-type jets, clearly
visible within filamentary lobes (Swain et al. 1998). The jets in
3C~353, constituting about $1\%$ of the entire source luminosity, are
well collimated trains of knots with an average width $\approx
4\arcsec$ and jet-counterjet radio brightness asymmetry $\approx 2$
(Swain 1996). Both total and polarized intensity profiles across the
jets indicate that the bulk of the jet radio emission is produced at
the jet edges, and so presumably within a boundary shear layer (Swain
1996, Swain et al. 1998). Since these jets are among only a very few
FR~II jets wide enough to be resolved in X-rays, we planned and
conducted a deep {\it Chandra} observation of 3C~353, in order to
investigate the multiwavelength structure of powerful FR~II outflows.
We report on the results of these observations in the present paper.

\begin{figure}
\begin{center}
\includegraphics[angle=0,scale=0.43]{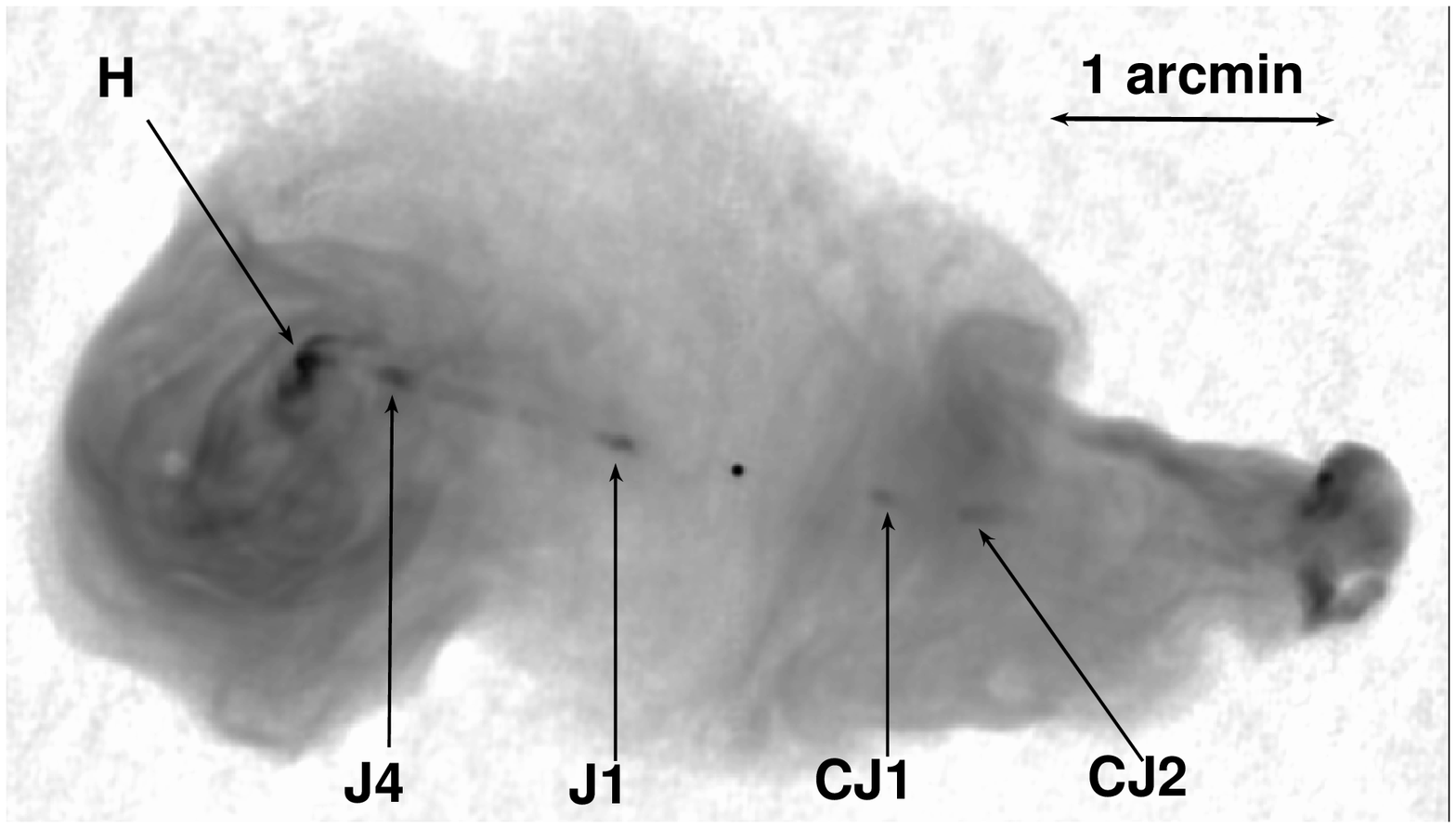}
\includegraphics[angle=0,scale=0.43]{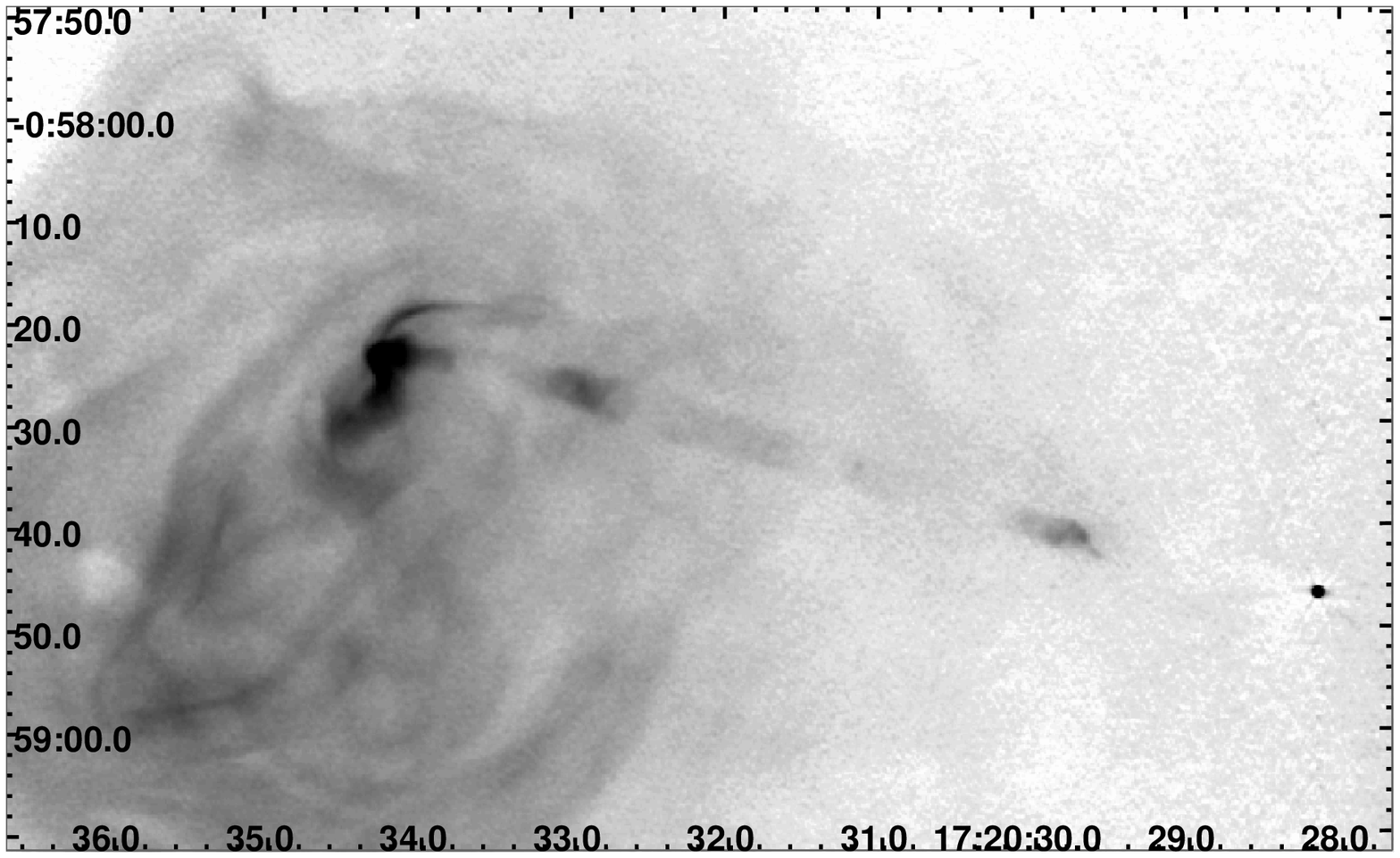}
\caption{Above: $1.4$\,GHz VLA image of 3C~353 at  
$1.3''$ resolution (adapted from Swain et al. 1998). 
Below: a zoom in on the eastern jet ($0.44''$ resolution at 
$8.4$\,GHz) shows that the jet is well resolved, and clearly 
extended compared to the unresolved nucleus.
}\label{fig:rad_img}
\end{center}
\end{figure}

\begin{table*}
\small{
  \caption{X-ray sources detected with \textsc{WAVEDETECT} along the 3C~353 jet.}
\label{tab:wavdetect}
  \begin{center}
    \begin{tabular}{lllcccc}
    \tableline
    X-ray & associated & nearby & RA (J2000)$^a$ & DEC (J2000)$^a$ & Net counts$^b$ & Significance ($\sigma$)$^d$\\
    jet region & X-ray features & radio knot & & & & \\
    \tableline\tableline
    EJ1 & E21/E23 & J1 & 17 20 29.7 & $-$00 58 41.6 & 46.9$\pm$7.2 & 13.7 \\
    EJ2 & E70/E73 & J4 & 17 20 32.8 & $-$00 58 26.6 & 12.0$\pm$4.0 &  3.8 \\
    EJ3 & E88 & H & 17 20 33.8 & $-$00 58 23.7 & 32.2$\pm$6.3 &  8.2 \\
    WJ1 & W47 & CJ2 & 17 20 25.0 & $-$00 58 55.1 & 32.0$\pm$6.0 & 10.0 \\
    WJ2 & W120a/b & ... & 17 20 20.3 & $-$00 59 09.9 & 12.3$\pm$3.7 &  4.8 \\
   \tableline
    \end{tabular}
   \tablenotetext{}{Note: all the errors are $1\sigma$.}
   \tablenotetext{a}{Coordinate center of the detected X-ray sources.}
   \tablenotetext{b}{Net photon counts given by the \textsc{wavdetect} 
   source detection algorithm.}
   \tablenotetext{c}{Statistical significance of detected jet features.}
   \end{center}
}
\end{table*}

A complication in X-ray studies of 3C~353 is the X-ray emission from
its large-scale environment. 3C~353 is situated at the edge of the
cluster Zw~1718.1-0108, which is a complex and dynamic system captured
at the moment of an on-going merger (Iwasawa et al. 2000). Based on a
$40$\,ks {\it ASCA} observation, Iwasawa et al. reported that this
cluster is characterized by a relatively low X-ray luminosity, $\sim 5
\times 10^{43}$\,erg\,s$^{-1}$, high temperature, $kT = 3-5$\,keV, and
disturbed morphology, extended over $\sim 30\arcmin$. The small field
of view of {\it Chandra}, together with the roll angle of {\it
Chandra} during our observations, prevent us from carrying out
detailed studies of the interaction between the radio source and the
cluster. Therefore, in this paper we focus solely on the
analysis of the non-thermal emission from the nucleus, jets, hotspots
and lobes. Analysis of the X-ray emission from the cluster and lobes
based on recent {\it XMM-Newton} observations is presented in Goodger
et al. (2008: hereafter G08); they found slightly different
temperatures for the southern and northern parts of the cluster, which
supported the idea that these two components were originally separate
features that are now undergoing a smooth merger, and also showed that
the X-ray emission detected from the lobes of 3C~353 was non-thermal
in origin, being consistent with the IC/CMB model prediction for the
lobes if the magnetic field strength is slightly below the
equipartition value, $U_{\rm B} \lesssim U_{\rm e}$.

The present paper is organized as follows. In
\S\ref{sec:observations}, we describe the {\it Chandra} observations
we performed and the data reduction process, and give a brief overview
of the X-ray image we have obtained. A detailed analysis of the jet
structures based on the X-ray/radio maps is presented in
\S\ref{sec:spatial}, and spectral analyses of the nucleus, jets, and
lobes are given in \S\ref{sec:spectral}. In \S\ref{sec:discussion}, we
discuss our findings in the context of various jet emission/particle
acceleration models. The final conclusions are presented in
\S\ref{sec:conclusion}. Throughout this paper we adopt a modern
cosmology with $\Omega_{\rm m} = 0.27$, $\Omega_{\rm \Lambda} = 0.73$
and $H_0 = 71$\,km\,s$^{-1}$\,Mpc$^{-1}$, leading to a luminosity
distance of $d_{\rm L}$ = 131.6\,Mpc and a conversion scale of
$0.60$\,kpc$/''$ for the 3C~353 redshift, $z = 0.0304$.

\section{Observations and Data Reduction}
\label{sec:observations}

Before going into the details of the {\it Chandra} X-ray data, let us first
remind the reader of the appearance of the jet of 3C~353 in high
resolution radio maps. Figure~\ref{fig:rad_img} shows NRAO Very Large
Array (VLA) images
of 3C~353 produced by Swain (1996). The upper panel is a deep-cleaned
L-band ($1.4$\,GHz) image with a resolution (FWHM of restoring convolving
Gaussian) of $1.3''$, similar to that of the {\it Chandra} X-ray
image. The lower panel shows a zoom in on the eastern jet region taken
at $8.4$\,GHz (X-band; $0.44''$ resolution). Full details of the data
reduction are described by Swain (1996). Note that the jet is well
resolved and clearly extended in the transverse direction, as compared to the
unresolved nucleus. Our goal in this section is to find out how the
{\it Chandra} X-ray image differs from, or agrees with, these radio 
images and provide one-to-one identification of various jet structures.
Although 3C353 is one of the brightest 3C sources, 
on the parsec-scale it is not as noteworthy. 
A \lq\lq snap-shot" Very Long Baseline Array (VLBA) 
observation at 4.8 GHz only detected the 
source on the shortest baselines. Fomalont et al.\ (2000) conclude 
the core of 3C353 is only $\sim$150 mJy, with a size of $\sim$15 
milli-arcseconds.

3C~353 was observed with {\it Chandra} in July 2007 with a total
(requested) duration of $90$\,ks (SeqNum 701549; ObsID 7886, 8565). The
source was observed using the standard ACIS-S configuration, where the
back-illuminated S3 chip was used to obtain maximal sensitivity to
soft photons. After deleting a high background interval at the end of
ObsID 7886 (Jul 04 18:22:16$-$18:30:36 UT), the exposure times
for the two observations were $70.4$\,ks (ObsID 7886) and
$17.9$\,ks (ObsID 8565), respectively. The active nucleus was at
the aim point, resulting in both sub-arcsecond resolution and
good spectral sensitivity for the whole jet (including knots,
hotspots and lobes). The roll angle was constrained
to preclude the possibility that any readout streak from
the nucleus ($F_{\rm 2-10\,keV} \sim 10^{-12}$\,erg\,
cm$^{-2}$\,s$^{-1}$; Iwasawa et al. 2000; G08;
see also \S\ref{sec:nucleus}) could affect the imaging of the jet. Although
the imaging capability of the High Resolution Mirror Assembly (HRMA)
degrades substantially ($\geq 1''$) for sources more than $3'$
off-axis\footnote{See {\it ``Proposers' Observatory Guide v.10''} at
\texttt{http://asc.harvard.edu/proposer/POG/}.}, the jet in 3C~353
extends over about $\pm 2'$ in length, so that the $50\,\%$
encircled energy for a point source is contained within a
diameter of $\sim 1''$ all along the 3C~353
outflows\footnote{See, in this context, \S\ref{sec:jetwidth},
for a detailed simulation using MARX software.}.
The level-2 data were reprocessed using the CXCDS software CIAO~3.4
and CALDB~3.4.3.
We generated a clean data set by selecting the standard {\it ASCA}
grades (0, 2, 3, 4 and 6) and energy band $0.4-8.0$\,keV.

\begin{figure}
\begin{center}
\includegraphics[angle=0,scale=0.43]{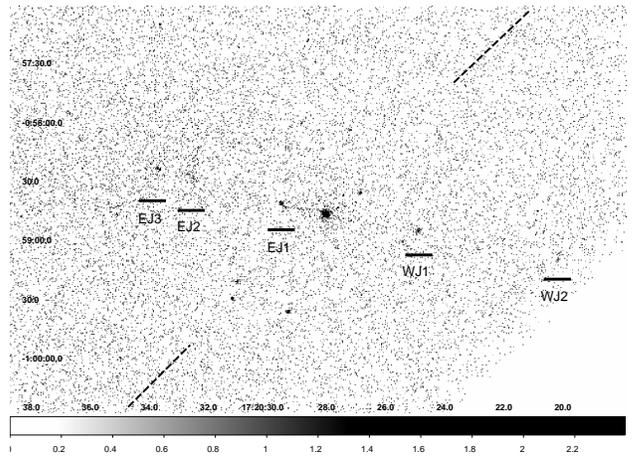}
\caption{A raw X-ray image of 3C~353 obtained with {\it Chandra} 
for in the $0.4-8.0$\,keV energy band. Five jet regions 
detected with \textsc{wavdetect} on angular scales of 
$0.5''-8''$ (wavelet radii of 1, 2, 4, 8, and 16 pixels) 
are labelled as EJ1, EJ2, EJ3, WJ1, and WJ2, respectively
(see also Table\ref{tab:wavdetect}).
Dashed lines show the direction of readout streak of the 
ACIS-S3 CCD chip, showing that the bright nucleus does not  
affect measurements of the jet features.  
}\label{fig:raw_X}
\end{center}
\end{figure}

A co-added, raw image of the 3C~353 X-ray jet in the $0.4-8.0$\,keV
bandpass is shown in Figure~\ref{fig:raw_X}. The dashed lines show the
direction of the readout streak of the ACIS-S3 CCD chip, confirming that
the bright nucleus does not affect our observations of the jet.
The {\it Chandra} image shows that the X-ray emission
consists of a bright nucleus,
jet-knots, hotspots and diffuse emission possibly associated with
the East radio lobe. Using a wavelet decomposition source-detection
algorithm (\textsc{wavdetect}; Vikhlinin et al. 1995) to detect emission
enhancements on angular scales of $0.5''-8''$ (wavelet radii of 1, 2,
4, 8, and 16 pixels) with more than $3\sigma$ significance,
we find a total of 31 distinct X-ray sources or
enhancements with a signal threshold $10^{-6}$ (the threshold for
identifying a pixel as belonging to a source\footnote{See \texttt{http://cxc.harvard.edu/ciao3.4/download/doc/detect\_manual/.}})
in the analysis region on the ACIS-S3 chip\footnote{A square region
with physical (detector) coordinates 3500$\le$DETX$\le$4500 and
3600$\le$DETY$\le$4600, where the nucleus of 3C~353 is at (4086.7, 4131.6).}.
Among these, the brightest one is the nucleus of 3C~353,
but five sources are closely aligned with radio-knots in 3C~353 as marked
in Figure~\ref{fig:raw_X} (denoted as EJ1, EJ2, EJ3, WJ1, WJ2).
The statistical significance of these features are 13.7, 3.8, 8.2,
10.0, and 4.8 $\sigma$, respectively (see Table~\ref{tab:wavdetect}).
Most strikingly, the X-ray image clearly shows not only
the East (main) jet-knots, but also a bright knot in the West
counterjet (WJ1), that seems to be identified with the CJ2 radio knot
as given in Swain et al. (1998). Also X-ray emission near the
West hotspot region is detected in the image (WJ2).

Since there are some X-ray point sources in the field, it is important
to ask whether the the counterjet feature (WJ1) detected in X-rays
could just be a chance coincidence with a background or foreground
object. Of the 26 sources without radio counterparts in the analysis
region, 13 were coincident with optical stars in our Galaxy listed in
the USNO-B1.0 catalogue (Monetet al. 2003)\footnote{See also
\texttt{http://www.nofs.navy.mil/data/fchpix/} for a Web version of
the catalogue.}, and 13 were unidentified or newly found in this {\it
Chandra} X-ray observation. We also checked the NASA/IPAC
Extragalactic Database\footnote{NED; see
\texttt{http://nedwww.ipac.caltech.edu/}.} and found four galaxies
within a $5'$ radius of 3C~353, but none of these was detected in our
{\it Chandra} X-ray image. This leads us to conservatively estimate a
probability that the X-ray source detected in the vicinity of CJ2
(within $3''$ radius) is just a chance coincidence (i.e., not related
to the jet) is $\le 0.4\,\%$. An even stronger argument for WJ1 being
different from the background point sources is that it is well
resolved and broader than the point spread function, as we will
show in \S\ref{sec:jetwidth} (see also Figure~\ref{fig:trans}).

\begin{figure}
\begin{center}
\includegraphics[angle=0,scale=0.45]{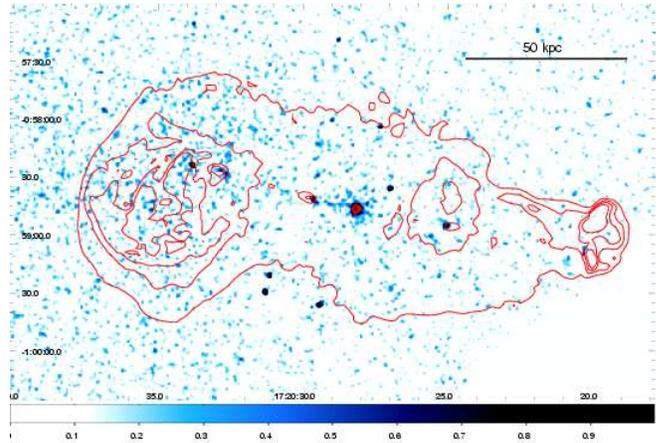}
\caption{An exposure corrected X-ray image of 3C~353 ({\it Chandra} 
ACIS-S3: $0.4-8.0$\,keV) smoothed with a two-dimensional Gaussian function 
with $\sigma = 1.5$ pixels using DS9, 
overlaid with radio contours (VLA, $1.4$\,GHz). 
The contour levels are $1.2$, $4.6$, $8.1$, and $11.5$\,mJy\,beam$^{-1}$. 
}\label{fig:X_img}
\end{center}
\end{figure}

\begin{figure}
\begin{center}
\includegraphics[angle=0,scale=0.45]{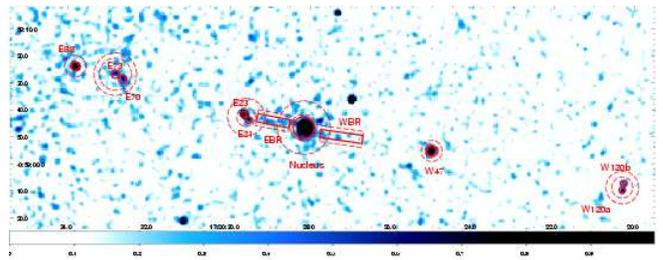}
\caption{A {\it Chandra} image of the central region of 3C~353,
smoothed with a two-dimensional Gaussian function 
with $\sigma = 1.5$ pixels using DS9 (0.4$-$8.0 keV).
The areas indicated by solid lines
mark regions from which X-ray spectra were extracted and
modeled with a power-law function plus Galactic absorption 
(Table~\ref{tab:wavdetect}).
Backgrounds were measured from the surrounding dashed annuli
or boxes.}\label{fig:region}
\end{center}
\end{figure}

Figure~\ref{fig:X_img} shows an exposure corrected image of 3C~353 in
the energy band $0.4-8.0$\,keV, with $1.4$\,GHz radio contours ($1.2$,
$4.6$, $8.1$, and $11.5$\,mJy\,beam$^{-1}$) overlaid. The X-ray image
has been smoothed with a two-dimensional Gaussian function with
$\sigma=1.5$ pixels (1 {\it Chandra} pixel is $0.492''$) using DS9
version 4.0. The nucleus is located in the center of the image, and
the jet extends to the East (main jet) and the West (counter jet)
directions. Figure~\ref{fig:region} shows the central region of
3C~353. The East jet (EJ1 and EJ2) shows rather complicated features
with multiple enhancements or peaks even inside each source region. We
therefore re-examined the source-detection algorithm by running
\textsc{wavdetect} with a single wavelet radius, changing its value
from 1 to 16 pixels. Interestingly, these jet features were repeatedly
detected at various wavelet radii, suggesting that there is relatively
narrowly peaked (i.e., point-source-like) emission embedded in an
extended component, and/or there is another peak of emission very
close to the primary peak. Furthermore, we note that the centroid of
each detected regions exhibits a gradual shift as we move to larger
wavelet radii, although the positions are in good agreement for
background point sources. Unfortunately, the photon statistics are not
good enough to fully describe the complexity of the detected jet
features. We therefore approximate the EJ1 and EJ2 regions below as a
composite of two emission peaks, as indicated in
Figure~\ref{fig:region} (denoted as E21/E23 and E70/73, respectively).
Similarly, the EJ3 and WJ1 regions are denoted as E88, W47 according
to the distance from the nucleus of 3C~353 (see
Table~\ref{tab:wavdetect} and Figure~\ref{fig:region}). A
weak but rather complicated X-ray feature (W120 a,b:
Figure~\ref{fig:region}) is also detected near the West (counterjet)
termination region, but its nature is not obvious because it is rather
close to the edge of the S3 CCD chip in the ACIS-S array. Net photon
counts from these jet-related structures, after subtracting the
background photons, are given in Table~\ref{tab:wavdetect}. In
addition to the bright nucleus and knot-like (jet or hotspot-related)
features, we can see weak diffuse emission that bridges the innermost
jet knots and the nucleus, and that is only visible in the X-ray image
(regions marked as EBR and WBR in Figure~\ref{fig:region}, detected at
the $\sim 3\sigma$ level). No counterpart to this structure was found
in the $1.4$\,GHz radio map.

\begin{figure}
\begin{center}
\includegraphics[angle=0,scale=0.45]{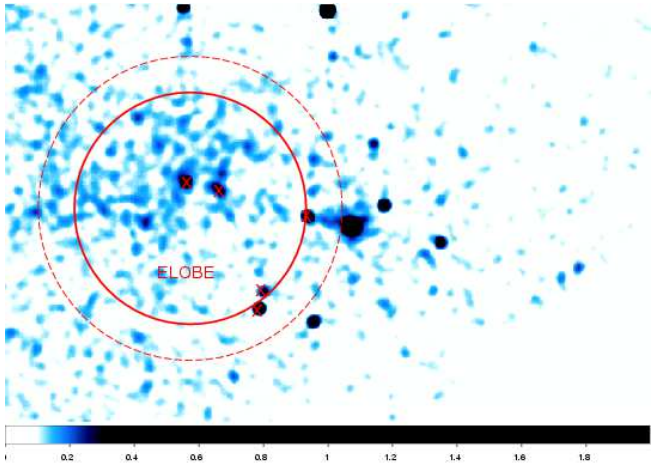}
\caption{A {\it Chandra} image of the lobe region of 3C~353, 
smoothed with a two-dimensional Gaussian function 
with $\sigma = 4$ pixels using DS9 (0.4$-$8.0 keV). 
The areas indicated by solid lines
mark regions from which spectra of lobe X-ray emission were extracted and
modeled with a power-law function plus Galactic absorption (Table~\ref{tab:spectra}).
The backgrounds was measured from the dashed annulus. Crosses show the 
positions of point sources (jet knots and background point 
sources) which were excluded when extracting the lobe 
spectrum.}\label{fig:lobe}
\end{center}
\end{figure}

Figure~\ref{fig:lobe} shows a more heavily smoothed image (smoothed
with a Gaussian with $\sigma = 4$ pixels) of 3C~353. Some excess
emission was found at the East lobe region, which corresponds to the
detection of this region by G08, but no emission was
detected exceeding the background level at the location of the West
radio lobe in our {\it Chandra} image. This is simply due to the
relatively poor photon statistics of {\it Chandra} compared to {\it
XMM-Newton}; the expected flux in the West lobe region is less than a
quarter of that of the east lobe region and hence is difficult to
detect with our {\it Chandra} exposure. Although the primary purpose
of this paper is to present the properties of the X-ray emission
associated with the 3C~353 jet, we will also briefly present the
analysis of the bright nucleus and East radio-lobe in later sections
of the paper in order to compare our results with those
obtained recently with {\it XMM-Newton} (G08).

\section{Image Analysis}
\label{sec:spatial}

\begin{figure}
\begin{center}
\includegraphics[angle=0,scale=0.4]{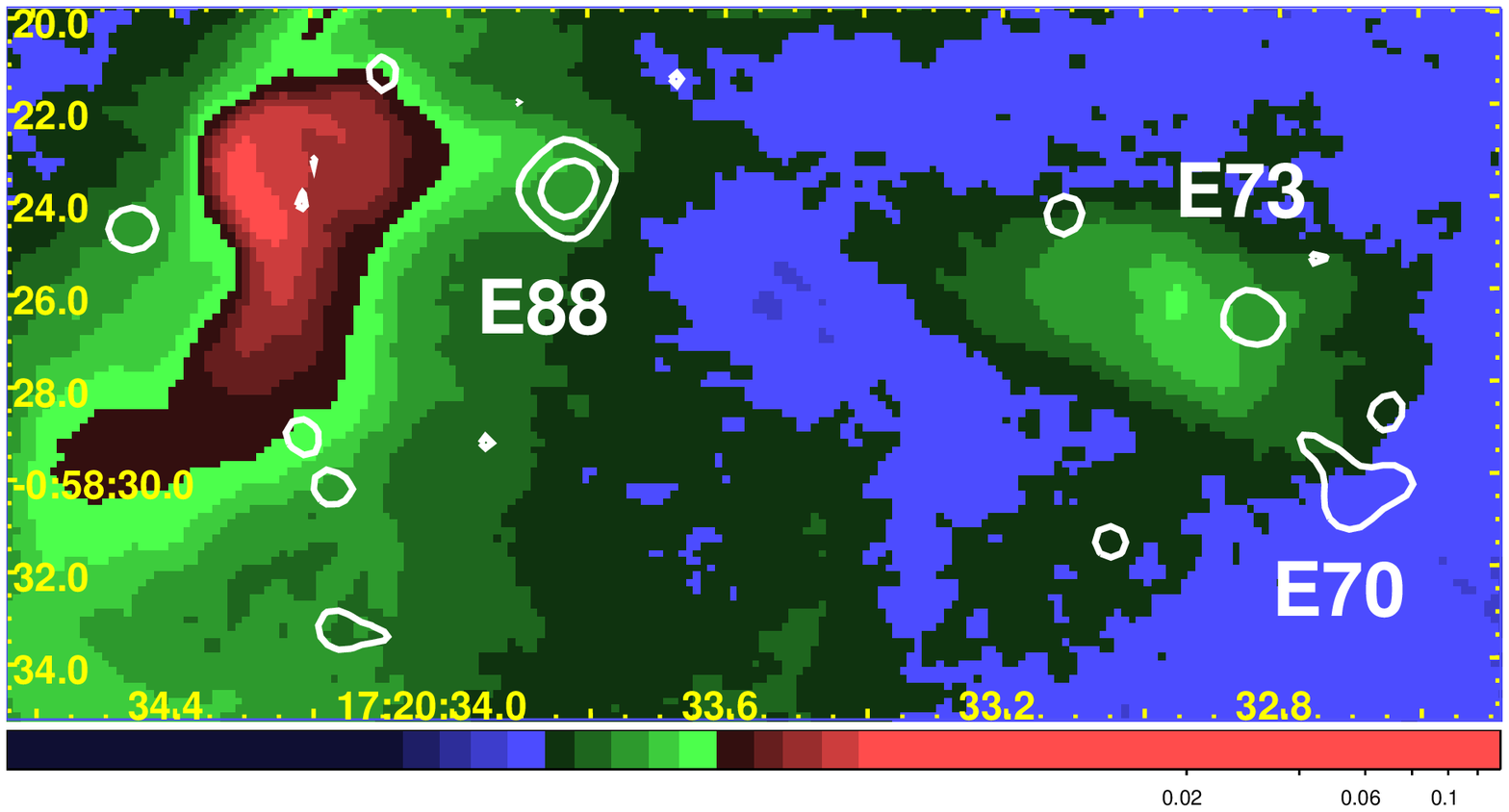}
\includegraphics[angle=0,scale=0.4]{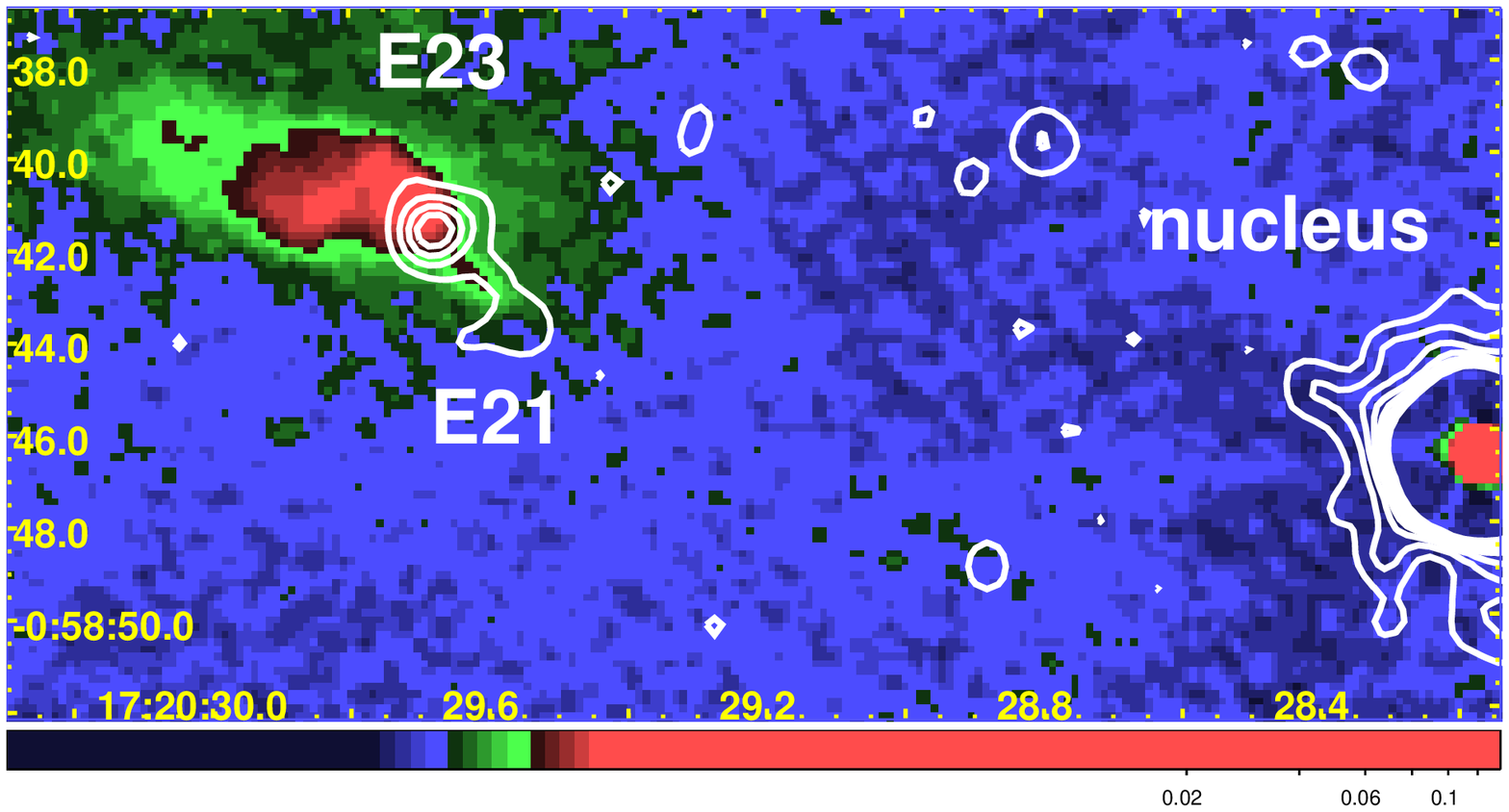}
\includegraphics[angle=0,scale=0.4]{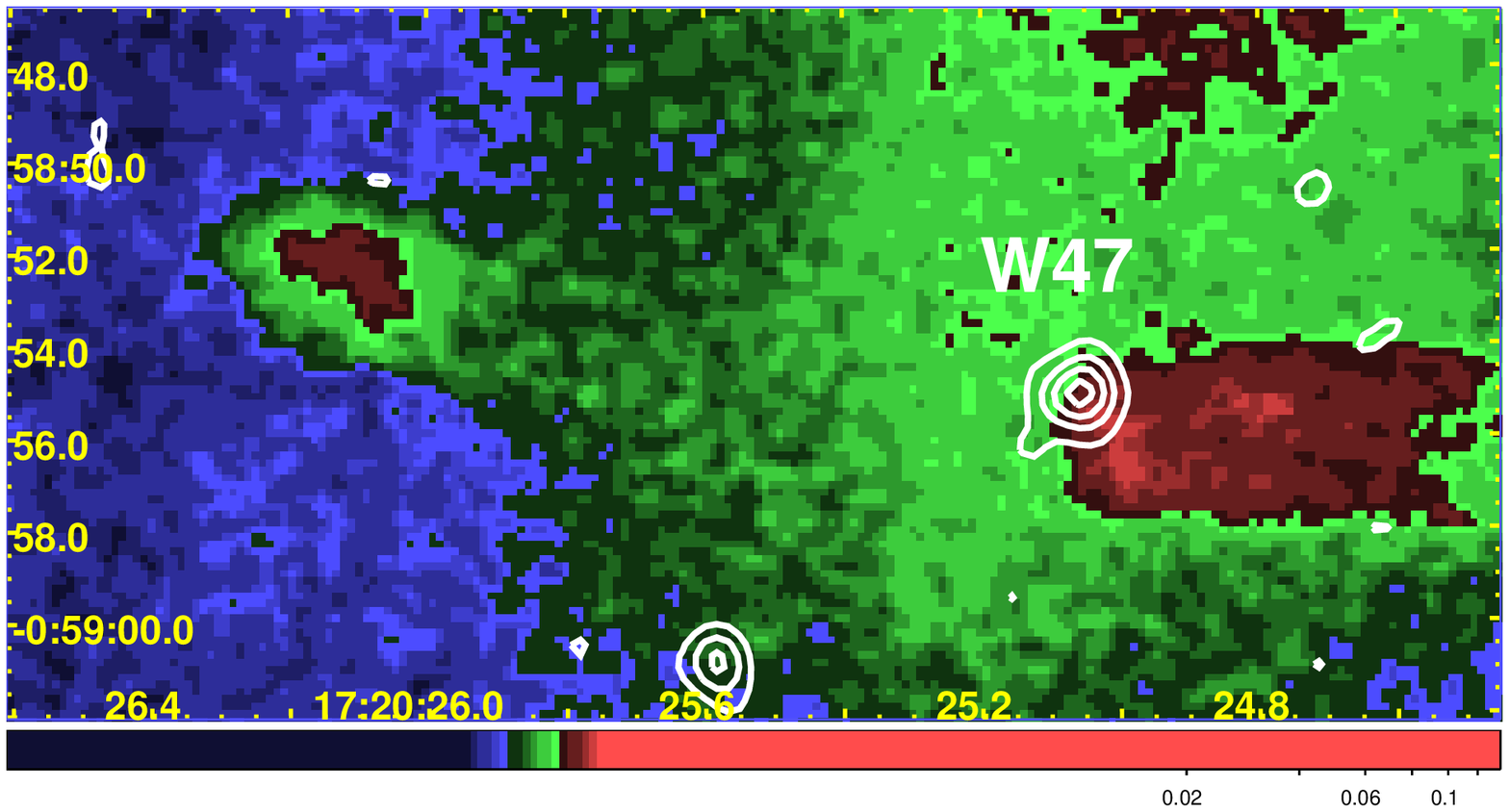}
\includegraphics[angle=0,scale=0.4]{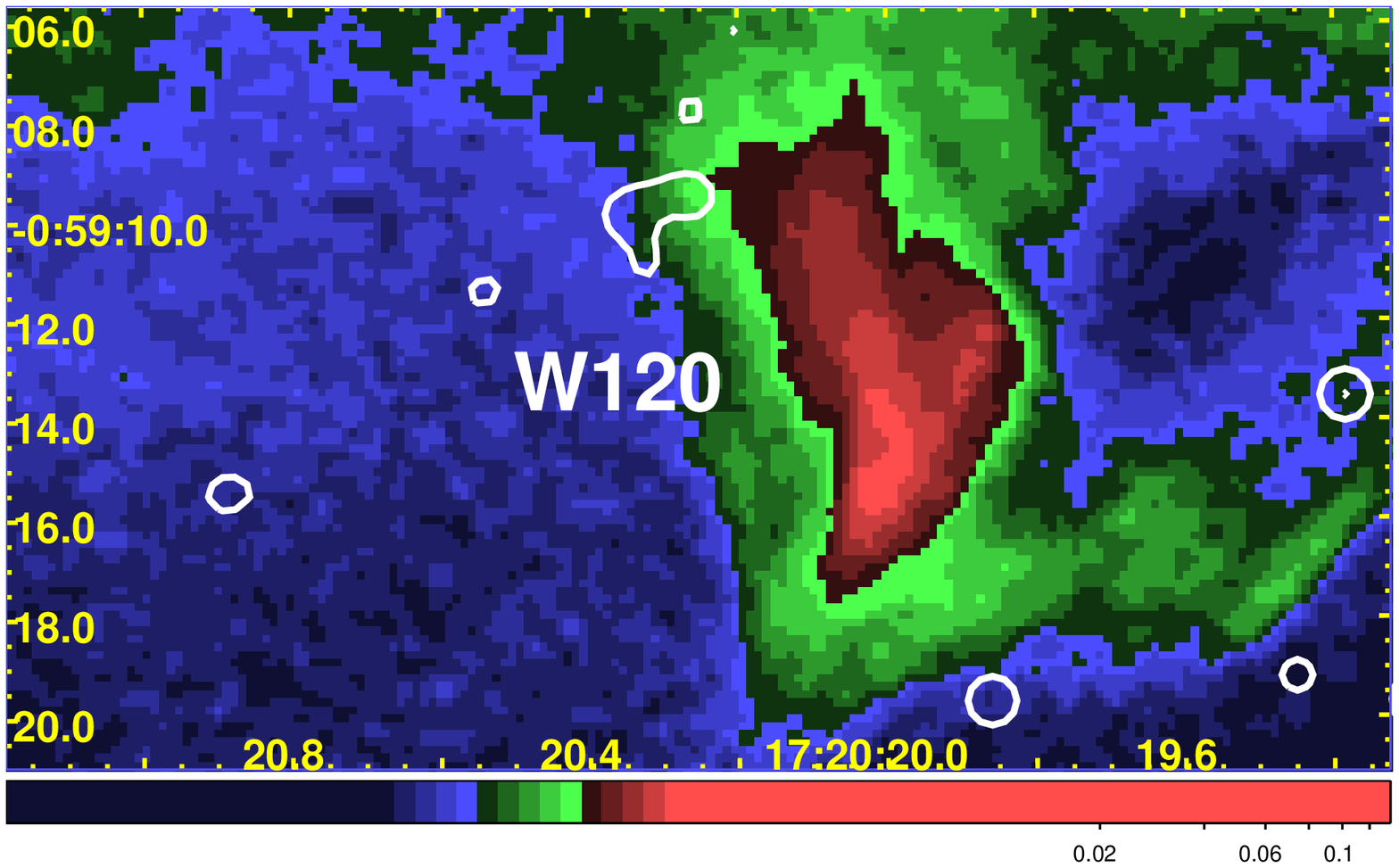}
\caption{Close-ups of the 3C~353 radio jets at $8.4$\,GHz with $0.4"$
 beam FWHM (colors),  overlaid with the X-ray contours produced from a
 smoothed (FWHM$=1"$) flux map between $0.8$ and $7$\,keV. 
The contours start at $1.5 \times 10^{-17}$\,erg\,cm$^{-2}$\,s$^{-1}$ 
per $0.246"$\,pixel, and increase by 
$1.5 \times 10^{-17}$\,erg\,cm$^{-2}$\,s$^{-1}$ up to $12 \times 
10^{-17}$\,erg\,cm$^{-2}$\,s$^{-1}$.}\label{fig:close-ups}
\end{center}
\end{figure}

In order to understand the origin of X-ray emission associated with
the 3C~353 jet knots, we first derive the intensity (photon counts)
profile of the jet both in the longitudinal and transverse directions,
which enables the direct comparison of X-rays from 3C~353 with the
high resolution VLA images ($1.4$\,GHz and $8.4$\,GHz: Swain et
al. 1998). This allows us to study in detail the positional offsets
between the radio and X-ray peaks, and differences in the jet width,
as measured perpendicular to the main jet axis. Details of the
radio/X-ray morphology for different jet features are shown again in
Figure~\ref{fig:close-ups} (see also Table~\ref{tab:wavdetect} for the
nomenclature of radio and X-ray jet knots).

\subsection{Longitudinal Jet Profile: Positional Offset}

\begin{figure}
\begin{center}
\includegraphics[angle=0,scale=0.50]{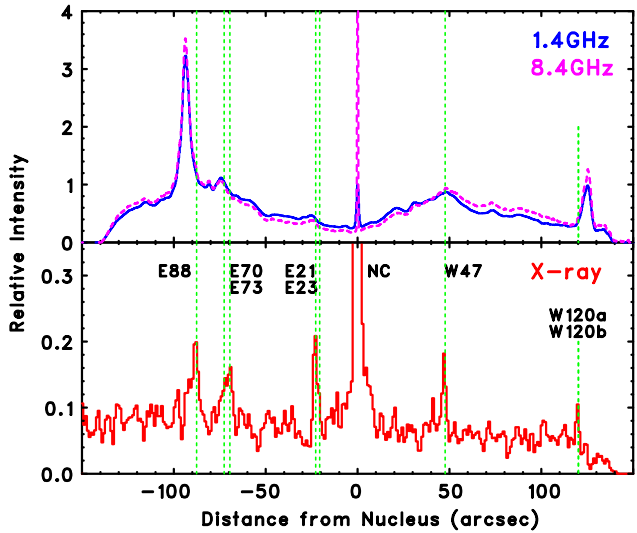}
\caption{Intensity profiles along the major axis of the lobes of 
3C~353. Above: the radio emission at $1.4$\,GHz (blue) and $8.4$\,GHz 
(magenta), Below: the X-rays in the $0.4-8.0$\,keV range. Intensities 
of radio/X-rays emissions are normalized to their nuclear fluxes 
($\times 1$ for $1.4$\,GHz, $\times 10$ for $8.4$\,GHz and X-rays).
The green dotted lines show the centers of jet knots and
hotspots, which are defined in Figure~\ref{fig:region}.
}\label{fig:rad_X_int}
\end{center}
\end{figure}

We first integrated the photon counts in a rectangular region of $\pm
12.8''$ width around the main jet axis. Then the longitudinal
intensity profiles were constructed from $x = -150''$ to $+150''$
(defined from the east to the west, where $x = 0$ represents the
position of the nucleus). Figure~\ref{fig:rad_X_int} shows the
longitudinal jet profiles thus produced in both radio and X-ray. This
profile confirms the detection of jet knots (E21, E23, E70, E73, W47),
of X-ray features preceding the radio terminal regions (E88, W120a,b),
and finally of excess emission in the East lobe regions. In the radio
profiles, the intensity gradually increases toward the terminal hotspots,
but the same trend is not so clearly observed in the X-ray intensity
profile.

\begin{figure}
\begin{center}
\includegraphics[angle=0,scale=.47]{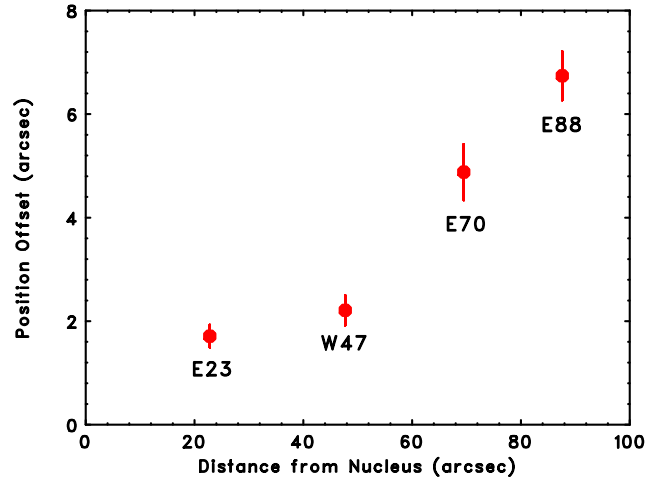}
\caption{Positional offset of jet knots and hotspots between 
radio and X-ray peaks in 3C~353, along the distance 
from the nucleus. 1 arcsec at the distance of 3C~353 corresponds to 0.60\,kpc. In this figure
the X-ray feature W120 was not included, since it lacks any obvious radio
counterpart.}\label{fig:offset}
\end{center}
\end{figure}

Figure~\ref{fig:rad_X_int} also makes it clear that there are
substantial offsets between X-ray and radio knots/hotspots, with the
centroids of the X-ray peaks being closer to the nucleus, a phenomenon
that has previously been observed in several other jet sources studied
with {\it Chandra} (see, e.g., Hardcastle et al. 2003; Siemiginowska
et al. 2002; but see also Siemiginowska et al. 2007 for the revised
analysis using new high S/N {\it Chandra} data).
Figure~\ref{fig:offset} shows the positional offset between radio and
X-ray peaks of the various jet knots and hotspot as a function of the
distance from the nucleus. To measure the offsets we have used the
$1.4$\,GHz map, since its resolution is $1.3''$, close to the
$\sim 1''$ resolution of the {\it Chandra} X-ray image. Clearly, the
measurements imply a smooth increase of offsets downstream along the
jet, which possibly saturates at a maximum value at the East hotspot 
(E88; $6.7 \pm 0.2''$, consistent with the value measured using {\it
  XMM-Newton} by G08).\footnote{The X-ray feature W120 is not considered
  to quantitatively estimate an offset, since it lacks any obvious 
radio counterpart.}
The maximum observed offset ($\sim 7''$) corresponds to large physical
distance ($\sim 4$\,kpc) at the distance of 3C~353. It is important
to mention in this context that there is no apparent mismatch or
discrepancy of this trend between the main jet (E23, E70, E88) and
counterjet (W47) suggesting the \emph{same} physical origin
for the X-ray production (see \S\ref{sec:origin} for detailed
discussion). Although
the X-ray emission from the hotspot region is coincident with the
start of the increase in the radio, the good (i.e., statistically
significant) detection of the X-rays is more than $\sim 7''$ upstream
from the peak of the radio emission from the East hotspot. There is
some weak radio emission closer to E88, so that it is plausible that
E88 may not be a direct counterpart of the East hotspot, but rather a
jet knot like the E23 and E73 X-ray features. This idea would be
supported by the relatively similar spectral energy distributions
(SEDs) of various jet knots, as we will see in \S\ref{sec:jet-spectra}. Similarly,
the X-ray emission detected near the West hotspot (W120a,b) may not be
related directly to the jet-termination structure in the radio. These
open issues cannot be addressed even using the excellent resolution of
both the {\it Chandra} and VLA images.

\subsection{Transverse Jet Profile: Jet Width}
\label{sec:jetwidth}

\begin{figure}
\begin{center}
\includegraphics[angle=0,scale=0.41]{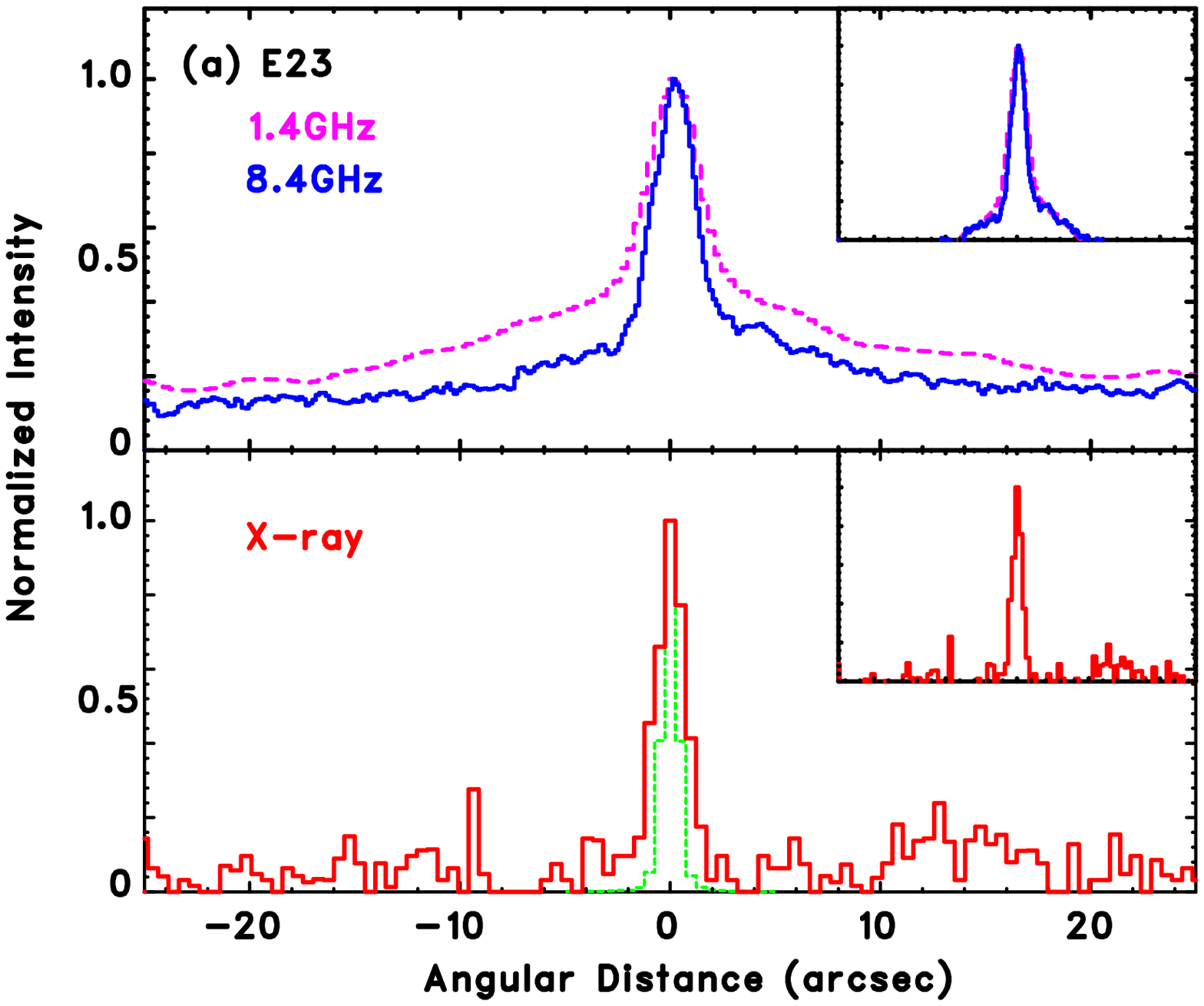}
\includegraphics[angle=0,scale=0.41]{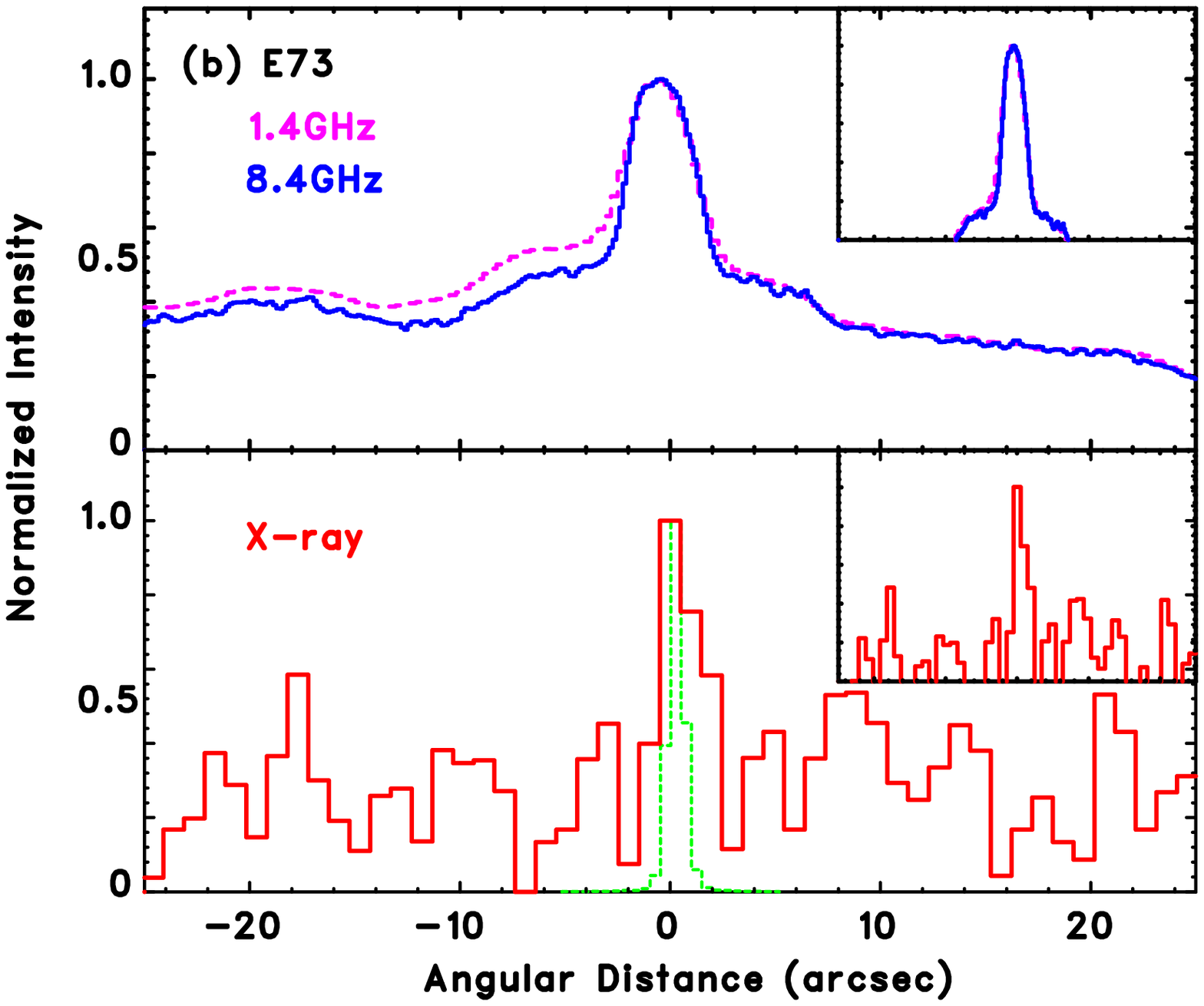}
\includegraphics[angle=0,scale=0.41]{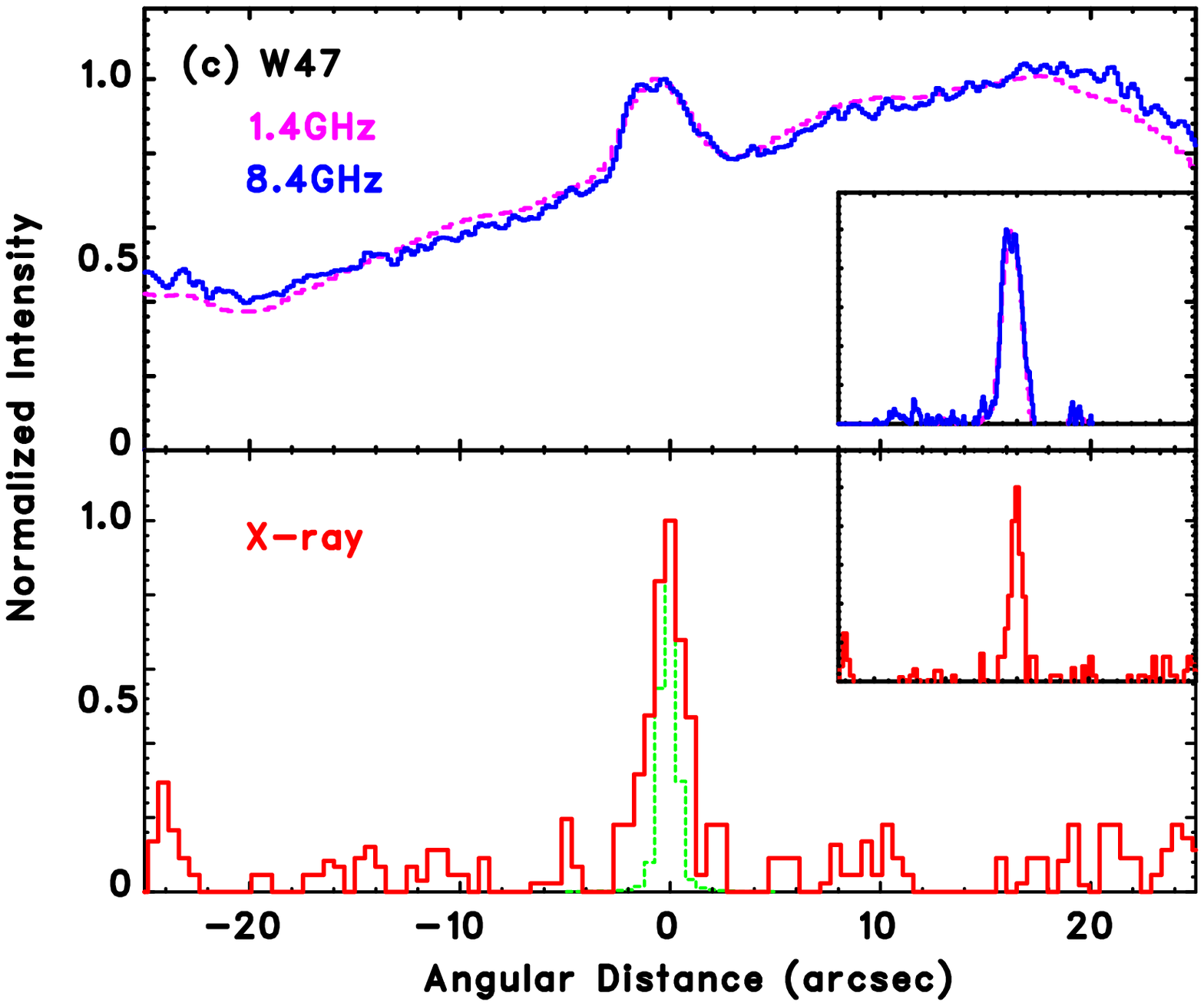}
\caption{(a) Comparison of transverse profile in the jet knot E23. 
Above: the radio emission at $1.4$\,GHz (blue) and $8.4$\,GHz 
(magenta). Below: the X-rays in the $0.4-8.0$\,keV range. A 
green dotted line shows the point spread function at the off-axis 
distance of the source simulated with MARX, by assuming a source spectrum with 
photon index $\Gamma = 1.7$ modified by Galactic absorption 
$N_{\rm H}^G = 1.6 \times 10^{21}$\,cm$^{-2}$.
In each panel, the insets to the upper right present the jet profiles
 after subtracting the smooth variation of the underlying lobe emission. Radio
resolutions are the same as in Figure~\ref{fig:rad_img}.
(b) Same as (a), but for the jet knot E73. (c) Same as (a), but for the jet knots W47.}\label{fig:trans}
\end{center}
\end{figure}

We next integrate the photon counts across the jet at certain
distances from the nucleus to construct transverse intensity profiles
of the jets. Here we concentrate on the profiles of the relatively bright
X-ray jet knots, E23, E73, and W47, perpendicular to the main jet
axis. We define the regions of integration to be small boxes ($3''$
along the main jet axis) centered on each jet knot, so as to reduce
the contamination from neighboring fainter jet knots, E21 and E70.
Figure~\ref{fig:trans} shows the transverse jet profiles thus produced
for (a) E23, (b) E73, and (c) W47 as compared with the corresponding
radio intensity profiles taken from the same regions. In each panel,
sub-panels show the profiles after subtracting
smooth variations of underlying lobe emission. The peak of the
intensity profile is normalized to unity in each panel for convenience
of comparison of the radio and X-ray profiles.

As discussed in Swain (1996) and Swain et al. (1998), the width of the
radio jet in all these figures is much broader than the resolution of
the radio map ($1.3''$ for $1.4$\,GHz and $0.44''$ for $8.4$\,GHz).
The transverse profile in the radio cannot be well represented by a
simple Gaussian function, but instead shows a flat-topped profile
which is especially clear in E73 and W47. For quantitative comparison,
the width of the jet, at which the intensity becomes half of the
maximum (FWHM), are $3.16''/2.66''$ (E23), $4.00''/3.77''$ (E73), and
$3.43''/3.32''$ (W47), where the widths are quoted for the $1.4$\,GHz
and $8.4$\,GHz radio maps respectively (see Table~\ref{tab:width}).
This flat-topped profile is characteristic of a situation in which the
bulk of the jet radio emission is produced within the jet boundary
(shear) layer, especially in the case of the counterjet knot W47
(corresponding to the radio feature CJ2 in Swain et al. 1998).

In contrast, the X-ray knots show no evidence for the 
flat-topped profiles seen in the radio images and are instead
well represented by a smooth Gaussian function. To investigate any possible 
broadening of the point spread function (PSF) at the position of 
each jet knot, we performed Monte-Carlo simulations\footnote{See
\texttt{http://space/mit.edu/ASC/MARX/marx\_4.0\_manual.pdf}.} 
using MARX 4.0. We modeled a point source with a power-law X-ray spectrum  
with a photon index $\Gamma = 1.7$, modified by Galactic 
absorption $N_{\rm H} = 1.6 \times 10^{21}$\,cm$^{-2}$, as 
indicated from modeling the observed X-ray spectra of the nucleus 
and jet knots (see \S\ref{sec:jet-spectra}). The resultant PSF is shown as a dotted 
green line in Figure~\ref{fig:trans}. The FWHM of the 
X-ray jet (E23, E73) and the counterjet (W47) are broader than 
that predicted by the PSF and hence the jet is \emph{well resolved} in
X-rays. 
In fact, the observed FWHM of the X-ray jets 
are $1.65'' \pm 0.27''$ (E23), $1.98'' \pm 0.48''$ (E73),  
and $1.79'' \pm 0.36''$ (W47), respectively, as compared to the 
PSF width of $0.93'' \pm 0.01''$ (Table~\ref{tab:width}). 
Although the difference between the radio and X-ray jet widths 
is still marginal (a $\gtrsim 4\sigma$ effect even for the most 
significant case, the jet knot W47), these results provide important 
information about the internal jet structures and the underlying jet 
physics, as we will discuss in \S\ref{sec:jet-physics}.

\begin{table}
\small{
  \caption{Width of 3C~353 jet knots as measured from radio/X-ray images.}
\label{tab:width}
  \begin{center}
    \begin{tabular}{lcccc}
    \tableline
    Name  & beams width$^a$ & $\Delta w_{\rm 1.4}^b$ & 
    $\Delta w_{\rm 8.4}^b$ & $\Delta w_{\rm X}^c$\\
     & $1.4$\,GHz/$8.4$\,GHz/X-ray & & & \\
     & [$''$] & [$''$] & [$''$] & [$''$]\\
    \tableline\tableline
    E23 & 1.3/0.44/0.92 &  3.16 & 2.66    &  1.65$\pm$0.27 \\
    E73 & 1.3/0.44/0.94 &  4.00 & 3.77    &  1.98$\pm$0.48 \\
    W47 & 1.3/0.44/0.93 &  3.43 & 3.32    &  1.79$\pm$0.36 \\
   \tableline
    \end{tabular}
   \tablenotetext{}{Note.: all the errors are $1\sigma$.}
   \tablenotetext{a}{FWHM width of the circular Gaussian beams which was used 
   to restore the $1.4$\,GHz/$8.4$\,GHz radio images; the X-ray width is 
   estimated from the Gaussian dispersion obtained from a   
   simulation using MARX at various offset angles (see \S\ref{sec:jetwidth} for
   detailed description of the X-ray simulation).}
   \tablenotetext{b}{FWHM $1.4$\,GHz/$8.4$\,GHz width of various jet knots measured
   by using the projected profiles perpendicular to the main jet axis from the 
   deconvolved radio images presented in Figure~\ref{fig:rad_img}.}
   \tablenotetext{c}{FWHM width of various jet knots measured 
   from the $0.4-8.0$\,keV {\it Chandra} image.}
   \end{center}
}
\end{table}

\section{Spectral Analysis}
\label{sec:spectral}

We next present detailed results from the spectral analysis of 3C~353's 
nucleus, jet knots and radio lobe. We find that the nucleus shows 
a complex X-ray spectrum with a heavily absorbed power-law component, while 
both the jets and radio lobe are well represented by a simple power law 
moderated by Galactic absorption. We find a smooth variation of 
radio ($1.4$\,GHz) to X-ray ($1$\,keV) flux density ratio as a function of 
distance from the nucleus for the jet. Finally, the spectral energy 
distributions of several bright jet knots are presented in order to
investigate the origin of the X-ray emission from the counterjet knot W47.

\subsection{Nucleus}
\label{sec:nucleus}

We extracted the spectrum of the nucleus from a circular region with a
radius of $4''$, while the background 
was estimated from a local background region with an annulus of an 
outer radius of $10''$ and inner radius of $5''$
(see also Figure~\ref{fig:region}). 
We obtained $2833 \pm 54$ net photons from the nucleus, combining data
from ObsIDs 7886 and 8565.
During our {\it Chandra} observations, no variability was detected for 
the nucleus on day-to-week scale with a constant X-ray count rate 
of $\simeq 0.03$\,cts\,s$^{-1}$ in the $0.4-8.0$\,keV energy band. 
For the spectral fitting, an accurate estimate of the Galactic 
absorption ($N_{\rm H}^G$) is obviously important. At this point, we note 
that the column density towards 3C~353 is uncertain, as has already been
pointed out in the literature. Iwasawa et al. (2000) adopted a value of 
$1.0 \times 10^{21}$\,cm$^{-2}$ based on the HI measurements 
of Dickey \& Lockman (1990). The Dickey \& Lockman value was questioned by G08, based on the Galactic dust measurements of Schlegel et al. 
(1998), which would provide a column density of a factor of 
$\sim 2$ higher ($\sim 2.6 \times 10^{21}$\,cm$^{-2}$). G08
argued that the true value is likely to lie somewhere between these two 
extremes; they used their new 
{\it XMM-Newton} data to \emph{directly} estimate a column density
from background sources
and found a weighted mean $N_{\rm H}^G = 
(1.64 \pm 0.07) \times 10^{21}$\,cm$^{-2}$. In this paper, we 
follow this `latest' estimate of the $N_{\rm H}^G$ value, 
fixed at $1.6 \times 10^{21}$\,cm$^{-2}$, in our spectral analysis.

\begin{table*}
\small{
  \caption{Results of the {\it Chandra} spectral fits to the nucleus in
 3C~353.}\label{tab:nucleus}
  \begin{center}
    \begin{tabular}{lllllllll}
    \tableline
    Model & $\Gamma_1$$^a$ & $F_{\rm 1\,keV}$$^b$  &  $F_{\rm
    0.5-5\,keV}$$^c$ & $N_{\rm H}$$^d$ & $\Gamma_2$$^e$  &  
    $F_{\rm 1\,keV}$$^b$ & $F_{\rm 0.5-5\,keV}$$^c$ & $\chi^2$ (dof)\\
     & & [nJy] & [$10^{-15}$\,cgs] & [$10^{22}$\,cm$^{-2}$] & & [nJy] & [$10^{-15}$\,cgs] & \\ 
    \tableline\tableline
    PL1 & $-$0.76$\pm$0.03 & 2.74$\pm$0.13 & 202$\pm$4 & ... &
     ... & ... & ... & 2.94(221) \\
    PL2 & ... & ... & ... & 5.34$^{+0.40}_{-0.38}$ & 1.47$\pm$0.13 &
    128$^{+32}_{-25}$ & 964$^{+13}_{-11}$  & 1.19(220) \\
    PL1+2 & 1.62$\pm$0.13 & 14.3$^{+4.8}_{-4.4}$ & 11.2$\pm$1.9 & 
    6.25$^{+0.40}_{-0.39}$ & 1.62$\pm$0.13 & 195$^{+55}_{-42}$ & 
    1140$^{+170}_{-140}$  & 1.02(219) \\
    \tableline
    \end{tabular}
   \tablecomments{Note: all the errors are $1\sigma$. Galactic
   absorption is fixed at $N_{\rm H}^{G} = 1.6 \times 10^{21}$\,cm$^2$
   (see G08). \\
$^a$: Photon index of power-law component 1 (modified by 
   Galactic $N_{\rm H}^{G}$ only). \\
$^b$: Absorption corrected flux   density measured at $1$\,keV, 
in units of nJy. \\
$^c$: Absorption    corrected flux in $0.5-5$\,keV, in units of 
   $10^{-15}$\,erg\,cm$^{-2}$\,s$^{-1}$. \\
$^d$: Intrinsic absorption    $N_{\rm H}$ at the distance of 3C~353 ($z = 0.0304$), in units of 
   $10^{22}$\,cm$^{-2}$. \\
$^e$: Photon index of power-law component-2   
(modified by intrinsic $N_{\rm H}$).}
   \end{center}
}
\end{table*}

\begin{figure}
\begin{center}
\includegraphics[angle=0,scale=.47]{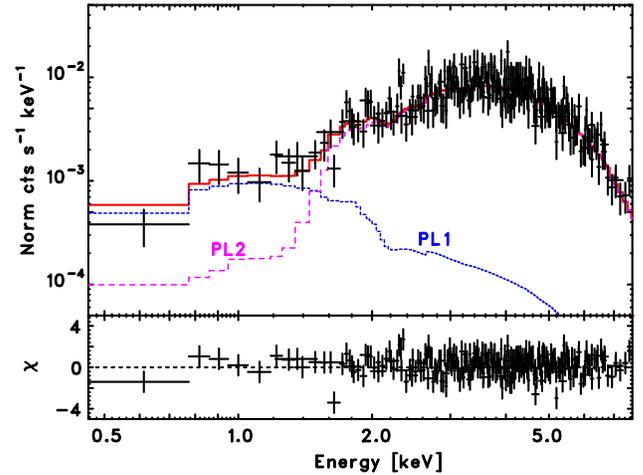}
\caption{Background-subtracted ACIS-S3 spectrum of the 
nucleus of 3C~353. Data points are plotted as crosses; the dotted and dashed lines represent the 
power-law component with Galactic absorption (PL1; Table~\ref{tab:nucleus}), and the second 
power-law component with redshifted intrinsic absorption (PL2), while
the solid line is their sum. The bottom 
panel shows the residuals for the best-fit double-power law models.}
\label{fig:specNC}
\end{center}
\end{figure}

We first attempted to fit the nuclear spectrum with a single power-law
function modified by Galactic absorption only. However, the fits were
very poor: $\chi^2 = 650$ for $221$ degrees of freedom (see
Table~\ref{tab:nucleus}). An attempt to fit with a thermal
bremsstrahlung model gave an even worse result, with $\chi^2 = 2254$
for $221$ degrees of freedom (dof). We therefore fitted a model by
adding an intrinsic absorption column density $N_{\rm H}$ at the
distance of the source ($z$ = 0.0304). The fitting statistic was
substantially improved, such that $\chi^2 = 261$ for $220$ dof, but
still unacceptable in the sense that $\chi^2$ probability is less than
$5\,\%$ ($P(\chi^2) = 3\,\%$). Hence we adopted a model consisting of
a power-law modified by $N_{\rm H}^G$ only, plus an another power-law
heavily absorbed by an intrinsic $N_{\rm H}$. We constrained the
photon indices of both power-law components to the same value, so as
to reduce the number of free parameters. This model represents the
observed X-ray spectra very well, with $\chi^2 = 223$ for $219$ dof
($P(\chi^2) = 41\,\%$). A summary of the spectral fitting parameters
are shown in Table\ref{tab:nucleus}, and the resultant best-fitting
spectra are presented in Figure~\ref{fig:specNC}.

The power-law indices determined by our {\it Chandra} observations are
consistent with those reported in G08 using the {\it
XMM-Newton} data. There are two minor differences in the spectral
fitting: (1) G08 determined both power-law indices
($\Gamma_1$ and $\Gamma_2$) independently when fitting the spectrum,
and (2) they added a thermal component ($kT$ = 1 keV) to
provide a good fit, as was required for some other FR~II sources in
the study of Hardcastle et al. (2006). Regarding the first point, our
constrained fitting is justified because G08
obtained nearly equal best-fitting parameters for both power-law
indices, $\Gamma_1 = 1.49 \pm 0.10$ and $\Gamma_2 = 1.33 \pm 0.30$,
which are very close to our best-fitting value of $1.62 \pm 0.13$
(Table~\ref{tab:nucleus}). Furthermore, our best-fitting value for the
intrinsic column density, $N_{\rm H} = (6.3 \pm 0.4) \times
10^{22}$\,cm$^{-2}$, is perfectly consistent with that determined using
{\it XMM-Newton} ($N_{\rm H} = (6.1 \pm 0.6) \times
10^{22}$\,cm$^{-2}$). Although thermal emission is included in the
G08 fit to the {\it XMM-Newton} data, it contributes less
than $\sim 1\,\%$ of flux in the energy range between $0.6$ and
$8$\,keV (see Figure~4 of G08) and in any case may
well have originated on scales larger than our {\it Chandra}
extraction region given the much larger {\it XMM-Newton} PSF. We
therefore do not regard the differences between the spectral models
used as in any way significant.

Using our model, we determine the absorption corrected luminosity in
the $2-10$\,keV band for the heavily absorbed nucleus (PL2) is $(2.43
\pm 0.08) \times 10^{42}$\,erg\,s$^{-1}$. This is only $\sim 15\,\%$
lower than that reported for the {\it XMM-Newton} observations during
the August 2006 and February 2007 observations ($2.82 \times
10^{42}$\,erg\,s$^{-1}$), suggesting only minor variability of the
nucleus on a timescale of a year. By using the {\it ASCA} data,
Iwasawa et al. (2000) reported a factor of two larger luminosity in
the same $2-10$\,keV band ($4.5 \times 10^{42}$\,erg\,s$^{-1}$;
corrected for modern cosmology with with $\Omega_{\rm m} = 0.27$,
$\Omega_{\rm \Lambda} = 0.73$ and $H_0 =
71$\,km\,s$^{-1}$\,Mpc$^{-1}$) in their observations taken a decade
earlier, in September 1996. We suspect that contamination from the
neighboring cluster Zw~1718.1-0108 may be partly the reason for this
diagreement, since the PSF of {\it ASCA} ($\sim 3'$) is much larger
than that of either {\it XMM-Newton} or {\it Chandra}, but it probably
also indicates some real long-term variability in the AGN X-ray
output.

\subsection{Jet and Hotspots}
\label{sec:jet-spectra}
We next carry out a spectral analysis of various structures 
(including jet knots and hotspots) related to the 3C~353 jet. 
The areas indicated by solid lines in Figure~\ref{fig:region} 
mark regions from which spectra of the X-ray emission were extracted, 
while background spectra were extracted using the dashed regions. Parameters 
of the source extraction regions (distance from the nucleus to the 
center of the region, and source radius), as well as the results of spectral fitting, 
are summarized in Table~\ref{tab:spectra}. Due to the limited photon 
statistics, we binned the X-ray spectrum into three energy bands for 
both observations (ObsIDs 7886 and 8565): $0.3-1$\,keV, $1-3$\,keV and 
$3-10$\,keV. A power-law function absorbed by 
the Galactic column density (fixed to $N_{\rm H} = 1.6 \times
10^{20}$\,cm$^{-2}$; see above) gives acceptable fits 
with $\chi^2$ probability $P(\chi^2) \ge 15\,\%$ for $4$ or $5$ dof. 
The spectral photon index was fixed at $\Gamma = 1.7$ to match 
the radio photon index, except for the relatively bright jet 
knots E23/W47 and hotspot E88. (The photon index
determined for these bright jet knots is consistent with $1.7$ within 
the $90\,\%$ confidence range.)

\begin{table*}
\small{
  \caption{Results of the {\it Chandra} spectral fits to the jet knots,
 hotspots and lobe in 3C~353.}\label{tab:spectra}
  \begin{center}
    \begin{tabular}{llllllll}
    \tableline
    Name & Distance$^a$ & Region$^b$ & Net counts$^c$ &
     $\Gamma$$^d$ & $F_{\rm 1\,keV}$$^e$  & $F_{\rm 0.5-5\,keV}$$^g$ & $\chi^2$ (dof) \\
     & [$''$] & [$''$] & & & [nJy] & [$10^{-15}$\,cgs] & \\
    \tableline\tableline
    & & & main jet  & & \\
    \tableline
    EBR & 11.9 & [12, 3] &19.0$\pm$6.6 & 1.7$^f$ & 
         0.24$\pm$0.09 & 1.56$\pm$0.61 & 0.86(5)  \\
    E21 & 20.6 & 1.2 &10.2$\pm$3.5 & 1.7$^f$ & 0.09$\pm$0.04 & 
        0.57$\pm$0.24 & 0.78(5) \\
    E23 & 22.8 & 1.2 &36.2$\pm$6.2 & 1.81$^{+0.29}_{-0.27}$ & 0.42$\pm$0.09 & 
        2.60$\pm$0.47 & 0.84(4)\\
    E70 & 69.5 & 1.5 & 11.8$\pm$4.0 & 1.7$^f$ & 0.12$\pm$0.04 & 
        0.80$\pm$0.29 & 0.63(5)\\
    E73 & 72.6 & 1.5 & 9.8$\pm$3.7 & 1.7$^f$ & 0.08$\pm$0.04 & 
        0.50$\pm$0.24 & 0.69(5)\\
    E88 & 87.6 & 1.5 & 25.8$\pm$5.7 & 1.79$^{+0.63}_{-0.56}$ &
     0.24$\pm$0.08 & 1.46$^{+0.43}_{-0.41}$ & 1.64(4) \\
    ELOBE$^h$ & 83.6 & 60 & 620$\pm$71 & 2.17$^{+0.43}_{-0.39}$ &
     8.82$^{+1.55}_{-1.59}$ &  45.6$^{+8.1}_{-7.8}$ & 1.32(12)\\
    \tableline
    & & & counter jet  & & \\
    \tableline
    WBR & 14.1 & [16, 3] &20.0$\pm$6.3 & 1.7$^f$ &
     0.24$\pm$0.08 &  1.59$\pm$0.54 &  0.74(4) \\
    W47 & 47.7 & 2.0 &37.9$\pm$6.6 & 1.22$\pm$0.33 & 0.31$\pm$0.09 
        & 2.88$^{+0.55}_{-0.59}$ & 1.08(4)\\
    W120a$^i$ &120.3 & 1.2 &10.2$\pm$3.5 & 1.7$^f$ & 0.09$\pm$0.04
        & 0.59$\pm$0.23 & 0.54(5)\\
    W120b$^i$ &120.1 & 1.2 &7.2$\pm$3.0 & 1.7$^f$ & 0.07$\pm$0.03
        & 0.43$\pm$0.23 & 0.78(5) \\
   \tableline
    \end{tabular}
   \tablecomments{Note: all the errors are $1\sigma$. Galactic
   absorption is fixed at $N_{\rm H}^{G} = 1.6 \times 10^{21}$\,cm$^2$ 
   (see G08).\\
   $^a$:Distance of the X-ray feature from the nucleus.\\
   $^b$:Sizes of the source extraction regions; either the length and 
        width of a box (EBR, WBR) or the radius of a circle 
        (E21, E23, E70, E73, E88, ELOBE, W47, W120a, and W120b). \\
   $^c$: Net photon counts after subtracting the background photon counts.\\
   $^d$: Differential X-ray photon index.\\
   $^e$: Absorption corrected flux density measured at $1$\,keV, 
         in units of nJy.\\
   $^f$: Parameters were fixed to these values rather than determined
   from the data.\\
   $^g$: Absorption corrected flux in $0.5-5$\,keV, 
         in units of $10^{-15}$\,erg\,cm$^{-2}$\,s$^{-1}$.\\
   $^h$: The East lobe region defined here is exactly the same 
  as used by G08 for the {\it XMM-Newton} analysis, except that we have 
  masked the E70, E73, and E88 regions for our analysis.\\
  $^i$: Note that the detected feature is close to the
   CCD (ACIS-S3) edge (see Figure~\ref{fig:raw_X}).}
   \end{center}
}
\end{table*}

\begin{figure}
\begin{center}
\includegraphics[angle=0,scale=.47]{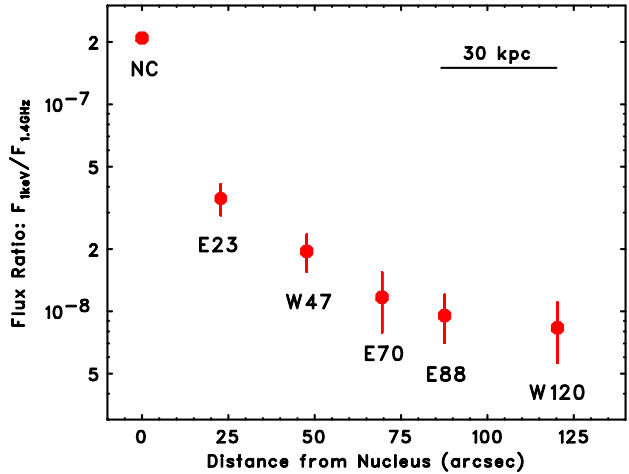}
\caption{Variation of X-ray ($1$\,keV) to radio ($1.4$\,GHz) flux density 
ratio along the major axis of 3C~353. Each plots involve the regions 
defined in Figure~\ref{fig:region}. Radio fluxes were extracted from exactly the
same jet volumes as X-ray fluxes (although the radio background regions were slightly 
shifted in order to avoid contamination from the bright radio structures downstream of
the outflows).}
\label{fig:ratio}
\end{center}
\end{figure}

Figure~\ref{fig:ratio} shows the variation of flux density ratio
measured between $1$\,keV and $1.4$\,GHz, $F_{\rm 1\,keV}/F_{\rm
1.4\,GHz}$, as a function of distance from the nucleus. Clearly the
X-ray-to-radio flux ratio decreases downstream along the jet, as is
often (though not always) observed in many other radio sources (see
\S\ref{sec:ratio-discussion}). Remarkably, this is true even 
for the counterjet.
Although X-ray counterjet detections were only made for W47 and,
possibly, W120a,b, we believe this to be due to the limited
sensitivity of {\it Chandra} even with our deep $90$\,ks exposure.
In fact, the $3 \sigma$ detection limit for this observation
corresponds to a flux limit $\sim 0.1$\,nJy. Thus, assuming the
jet-counterjet radio brightness asymmetry $\approx 2$ (Swain 1996) is
also valid for the X-ray counterparts, most of the X-ray emission from
the counterjet is not expected to be detected, with the exception of
the brightest counterjet knot W47. We note that the flux ratio of the
nucleus ($F_{\rm 1\,keV}/F_{\rm 1.4\,GHz}$; absorption corrected) is
much larger than that for the jet-related structures. This is not
surprising, given that the heavily absorbed component of the nuclear
X-ray emission is believed to be dominated by the accretion disk
corona, while the nuclear radio flux is supposed to be produced by the
unresolved, self-absorbed inner portions of the jets.

\begin{figure}
\begin{center}
\includegraphics[angle=0,scale=.47]{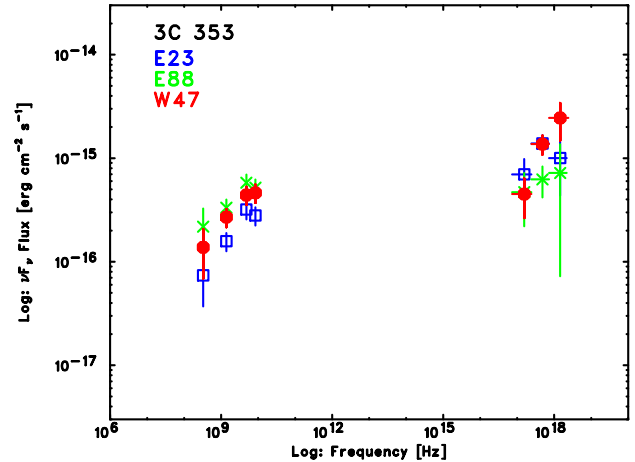}
\caption{Comparison of the spectral energy distribution of the jet knot 
(E23 in $blue$ and E73 in $green$) and the counterjet knot (W47; in 
$red$).}\label{fig:SED}
\end{center}
\end{figure}

Finally, we constructed the spectral energy distributions of the bright
jet knots E23, W47 and hotspot E88. For this, we took exactly same
source regions for the radio and X-ray as defined in
Figure~\ref{fig:region}. We derived multi-frequency radio spectral
information using radio maps obtained at four different frequencies
between $330$\,MHz and $8.4$\,GHz (adapted from Swain et al. 1998 for
$1.4$\,GHz, $8.4$\,GHz images, and G08 for $330$\,MHz
and $5$\,GHz radio maps). Since the resolutions of the radio maps are
different at these frequencies and the contamination from the lobe is
non-negligible, we made several trials assuming different choices of
background regions. We estimate that the uncertainty in the radio flux
measurements is typically $20-50\,\%$, but this is not important for
the discussion below. As before, the X-ray data were binned in three
energy bands to reduce the statistical uncertainty; $0.3-1$\,keV,
$1-3$\,keV, and $3-10$\,keV, respectively. Figure~\ref{fig:SED}
compares the spectral energy distributions (SEDs) thus produced, and
shows that the overall spectral features are remarkably similar to
each other, suggesting that \emph{the same} physical process is at
work for the X-ray production in the E23, E88 and W47 knots. Except in the
case of W47, the X-ray spectral index obtained with {\it Chandra} is
perfectly consistent with the radio synchrotron power-law slope (radio
spectral index $\alpha_{\rm R} \simeq 0.7$). The X-ray spectrum of W47
seems to be slightly harder, but this is only a $\lesssim 2\sigma$
effect. One may note, however, that the X-ray spectral points cannot
be connected smoothly with the extrapolation of the radio data,
whether we assume either a single or a broken power-law form for the
radio-to-X-ray continuum.

\subsection{Lobe}
\label{sec:lobe}

Finally we consider the X-ray emission from the East radio lobe, for
which the previous {\it XMM-Newton} observations confirm substantial
diffuse emission (G08). We do not consider the West radio lobe since
its emission was too weak to be detected within our {\it Chandra}
exposure (see \S\ref{sec:observations}). The areas indicated by the
solid line in Figure~\ref{fig:lobe} mark the region from which the
spectrum of the lobe X-ray emission was extracted (a circle of
$r=60''$ centered on RA=17:20:33.694, DEC =-00:58:37.77), but
excluding the inner jet knots (E21/23, E70/73), hotspots (E88) as well
as background point sources (marked as crosses) --- these were
approximated as circles of $r=1.5''$ each. The dotted line shows a
region of the background extraction with $r=78.7''$, excluding some 
background point sources. As discussed in detail by G08, the
background emission from the cluster environment is the primary
contributor to the uncertainty in the lobe flux density. We
therefore adopted exactly the same regions for both the source and
background extraction regions as those that were assumed for the {\it
XMM}-Newton data analysis, enabling a direct comparison of the
results.

A power-law model gave better fits than a thermal bremsstrahlung
model, with $\chi^2 = 15.8$ for $12$ dof, for the East lobe (see
Table~\ref{tab:spectra}). The best-fitting photon index was $\Gamma =
2.17^{+0.43}_{-0.39}$, which is consistent with that determined from
the {\it XMM-Newton} data analysis ($\Gamma = 1.9 \pm 0.4$; G08). The
measured $1$\,keV flux density of the east lobe is
$8.82^{+1.55}_{-1.59}$\,nJy, which is slightly smaller than, but
consistent within the $1\sigma$ error with, that determined from the
{\it XMM-Newton} analysis ($11.6 \pm 1.6$\,nJy; G08). The origin of
the possible (but small) discrepancy, however, is clear. In
G08, only the contamination from the East hotspot (E88) was excluded
to estimate the lobe X-ray emission because the remaining jet features
(E21, E23, E70, E73) were not resolved from the diffuse lobe emission
with the imaging capability of {\it XMM-Newton}. As seen in
Table~\ref{tab:spectra}, integration of $1$\,keV flux densities for
E21, E23, E70 and E73 gives $0.7$\,nJy in total, and hence we can
account for a substantial fraction of the difference between the {\it
Chandra} and {\it XMM-Newton} fluxes. Thus we conclude that the {\it
Chandra} data confirm the detection of non-thermal X-ray emission from
the East lobe of 3C~353 and the measurement of its parameters by G08.

\section{Discussion}
\label{sec:discussion}

In the previous sections, we have shown that X-ray emission has been
detected from the knots of both the jet and the counterjet of the
radio galaxy 3C~353. As argued below, this emission is most likely
non-thermal in origin. This is in fact the very first clear X-ray
detection of an extragalactic counterjet (or, more accurately, of
isolated counterjet knots) reported in the literature for a powerful
FR~II-type source (see the discussion in \S\ref{sec:intro}). We have
shown (i) that the X-ray knots seem to be narrower than their radio
counterparts, (ii) that the X-ray-to-radio flux ratio decreases downstream
along the jets, and finally (iii) that the intensity maxima of the X-ray
jet features are placed closer to the nucleus than the intensity
maxima of the nearest radio components (and that these offsets are
significant). Moreover, thanks to our deep $90$\,ks {\it Chandra}
exposure, the other components of 3C~353 have been successfully
detected as well, including the nucleus and the extended East lobe.
The nucleus and lobes will briefly be discussed further in
\S\ref{sec:frii} and \S\ref{sec:origin} respectively, but since our
major findings above are all related to the jets, the remainder of the
paper mostly focuses on the interpretation of the observed X-ray jet
emission. Before this, however, let us comment on the question of
whether 3C~353 is indeed representative of other quasar/FR~II jet
sources.
  
\subsection{Is It Indeed An FR~II Source?}
\label{sec:frii}
There are some reasons one might suspect that 3C~353 is not
representative of other powerful FR~II-type jets. First, 3C~353 is
located in the outskirts of a merging cluster, which is not a typical
environment for classical doubles or radio loud quasars. Second, one may
note the apparent lack of the X-ray signatures for the cavity inflated
in the cluster medium by the expanding lobes of 3C~353, like those
observed in, e.g., Cygnus~A (Smith et al. 2002). And third, the
particular morphology of the radio hotspots in the discussed object might
be thought to be unusual for an FR~II radio galaxy: the East
(main jet) hotspot is located in the middle --- and not at the edge, as
expected --- of the fat East lobe, while the counterjet terminates in
the elongated West lobe without forming a `classical' hotspot, but
instead splits into a mushroom-shaped vortex ring (see
Figure~\ref{fig:rad_img}). 
This type of jet morphology might rather be thought to resemble the `inner
hotspots', or `flaring regions' observed in Narrow Angle Tailed (NAT)
sources at the point where the substantially curved jets suddenly lose
their collimation but continue to propagate away from the nucleus rather
than terminating and forming backflow (see in this context Eilek et
al. 2002). Interestingly, NAT sources are typically (exclusively)
located in the outskirts of rich clusters. We therefore considered the
possibility that 3C~353 is a NAT radio galaxy observed in projection,
with the line of sight being almost in the jet bending plane.

The first reason to doubt the above interpretation is the very low
probability that the line of sight would be exactly in the plane in
which the NAT jet was bending. With any other orientation of the
source in the sky, one should expect to observe some gradual curvature
of the 3C~353 outflows, or at least some misalignment between the jet
and the counterjet axes. By contrast, the jets in 3C~353 are very straight all over their lengths. As for the jet
misalignment, we have carefully investigated the available radio maps,
and found only a small difference between the two axes: if a line is
drawn from the nucleus out to follow the West jet, and another from
the nucleus to follow the East jet, the position angles of the lines
differ by $5^{\circ}$. If, instead, we aim for similar size ellipses,
one fitting the hotspot region at the end of the counterjet and the
other (of the same size) fitting the brightest jet termination feature
of the main jet, the lines differ by $1^{\circ}$ only. The other
problem is that the jets in NAT sources are of the FR~I type, while
both the jets in 3C~353 --- extremely well collimated outflows with a
half-opening angle $\sim 0.5^{\circ}$ --- are clear examples of the
type of jets seen in FR~IIs (Swain 1996, Swain et al. 1998). Also, the
apparent lack of any signatures for the X-ray cavity inflated in the
ambient medium by the expanding lobes may be explained if 3C~353 is in
reality located in front of or behind the cluster mid-plane, as was
argued by G08 based on an analysis of pressure balance; even if the
source were in the plane of the cluster it is unlikely that a strong
cavity signature would be observed at the observed distance from the
cluster center. Finally, the recessed hotspot observed in the East
lobe of 3C~353 is not particularly unusual for classical doubles
(Mullin et al., submitted).

There are also many positive reasons for believing 3C~353 is a fairly
typical FR~II object, and here we briefly summarize them: (i) the total
radio power of 3C~353, $L_{\rm 1.4\,GHz} \sim 10^{26}$\,W\,Hz$^{-1}$, is
an order of magnitude above the borderline between FR~Is and FR~IIs
(Fanaroff \& Riley 1974); (ii) the spectral properties of the
X-ray nucleus in 3C~353, in particular a large intrinsic absorbing
column density $N_{\rm H}^G \sim 6 \times 10^{22}$\,cm$^{-2}$ (see
\S\ref{sec:nucleus} and G08), are typical to what is
observed in powerful narrow line radio galaxies, i.e. classical doubles
with narrow ionization lines\footnote{However, as noted by G08, the X-ray-to-radio flux ratio of the 3C~353 nucleus is much
lower than observed in other narrow line radio galaxies (NLRGs), and is
more consistent with what is seen in low-excitation radio galaxies
(LERGs), so that the nature of 3C 353's
nucleus in the X-ray remains ambiguous. Many LERGs exhibit
FR~II large-scale radio morphology.}; (iii) the position of 3C~353 in the
host (optical) luminosity--radio luminosity domain is consistent with
classifying 3C~353 as an FR~II source (Swain et al. 1998); (iv) the
detection of inverse-Compton X-rays from the lobes of 3C~353 is
consistent with what is typically seen in other classical doubles
(see \S\ref{sec:lobe} and G08). Thus, we conclude that 3C~353 
should be considered as representative of the class of powerful jet sources, 
including both radio-loud quasars and FR~II radio galaxies.

\subsection{Origin of the Observed X-ray Emission}
\label{sec:origin}
Although the constructed three-band X-ray spectra for the jet knots
(\S\ref{sec:jet-spectra}) are consistent with power-law emission, a
thermal origin of the detected X-ray features cannot be ruled
out by means of spectral analysis, because of the limited photon
statistics. In fact, one can fit the X-ray data for these knots with a
thermal model (bremsstrahlung radiation), although the derived gas
temperature is not well constrained. We therefore begin this section
by checking whether the global parameters implied by a thermal
interpretation are realistic.

The
bolometric bremsstrahlung luminosity of a hypothetical source with
volume $V$ and number densities of electrons and ions $n_{\rm e}$ and
$n_{\rm i}$, respectively, assumed to be characterized by a single
temperature $T$, is
\begin{equation}
L_{\rm ff} \approx 10^{-27} \, T^{1/2} \, \int \, n_{\rm e} \, n_{\rm i} \, dV \, ,
\end{equation}
\noindent
where all the quantities are expressed in cgs units (Blumenthal \&
Gould 1970). If we assume $kT \approx 5$\,keV (which is close to the
temperature of the Zw~1718.1-0108 cluster; Iwasawa et al. 2000;
G08), and consider a completely ionized thermal
hydrogen plasma ($n_{\rm e} = n_{\rm i}$) uniformly distributed within
the spherical volume $V = \pi \, r^3$ of the radius $r \approx 2''
\approx 1.2$\,kpc (as appropriate for, e.g., the W47 knot; see
Table~\ref{tab:spectra}), then we find $L_{\rm ff} \approx 10^{42} \,
n_{\rm e}^2$\,erg\,s$^{-1}$. The observed $0.5-5$\,keV
luminosity of the W47 knot is $L_{\rm X} \approx 6 \times
10^{39}$\,erg\,s$^{-1}$. This, in the framework of the thermal model,
would imply a number density of the thermal gas $n_{\rm e} >
0.07$\,cm$^{-3}$, which is much higher than what is expected for the
thermal gas density at the edge of the cluster. In addition, the
corresponding thermal pressure of the hypothetical X-ray emitting
cloud, $p_{\rm th} = n_{\rm e} \, kT > 6 \times
10^{-10}$\,erg\,cm$^{-3}$, would be about two orders of magnitude
higher than the non-thermal pressure of the lobe, $p_{\rm lobe} =
U_{\rm lobe} / 3 \approx 10^{-11}$\,erg\,cm$^{-3}$, as estimated
below. The presence of such a highly overpressured dense cloud of a
hot gas, with a total mass of $M_{\rm tot} = m_{\rm p} \, n_{\rm e} \,
V > 10^7 \, M_{\odot}$, within the lobes of a powerful radio galaxy and
close to the radio jet, seems highly unlikely. We also emphasize that
a clear discontinuity in the polarization properties of the jet, due
to the rotation measure associated with the cloud ($RM \approx 0.81 \,
n_{\rm e} \, (B_{\|} / {\rm \mu G}) \, (r / {\rm pc})$\,rad\,m$^{-2}$
$\gtrsim 100$\,rad\,m$^{-2}$, for an expected magnetic field component
parallel to the line of sight within the cloud $B_{\|} \gtrsim
1$\,$\mu$G), although not large, should have been observable in
the detailed analysis of the radio polarization maps presented by
Swain (1996) and Swain et al. (1998), but was not seen. Finally, we
note that decreasing the assumed gas temperature (in order to reduce
the pressure of the postulated X-ray emitting cloud) would cause the
derived gas number density, the total mass of the cloud, and the
rotation measures to increase still further. Therefore, we rule out a
thermal interpretation, and conclude that the X-ray emission of the
jet-related knots detected by {\it Chandra} is non-thermal in origin.

As can be seen in Figure~\ref{fig:SED}, the overall spectral energy
distributions of different knots in 3C~353 are similar to each other,
suggesting that {\it the same} physical process is responsible for
production of the observed X-ray emission from the jet and the
counterjet; thus, in what follows, we attempt to find a single
emission process that can explain all the observations. In addition,
as we have noted above, the X-ray spectra of all the knots are
inconsistent with simple extrapolations of the radio (synchrotron)
continua. In this situation, as discussed in \S\ref{sec:intro}, the
X-ray emission may be synchrotron emission from a second electron
population, or it may be some form of inverse-Compton emission. Since
the X-ray/radio morphology of the 3C~353 jet knots, discussed in more
detail in the next sections (in particular, the large positional
offsets between the X-ray and radio intensity maxima, as well as the
differences in the X-ray and radio jet widths), exclude any
significant contribution of the synchrotron self-Compton (SSC) process
to the observed X-ray radiative output, we do not consider the SSC
process further. Instead we focus our examination of inverse-Compton
processes on the IC/CMB model (\S\ref{sec:intro}). In core-dominated
quasars, the spectral behavior seen in the jet and counterjet knots of
3C~353 is often believed to be a manifestation of the IC/CMB component
dominating the observed X-ray emission, with corresponding high bulk
Lorentz factors (Sambruna et al. 2004, Marshall et al. 2005). We will
argue below that this cannot be the case in 3C~353, which
substantially weakens the case for adopting the model in analogous
FR~II/quasar jet sources.

We begin by briefly discussing the global energetics of the 3C~353 jets,
by utilizing our {\it Chandra} detection of the non-thermal X-ray
emission from the East radio lobe (see \S\ref{sec:lobe} and ELOBE in
Table 1), which is consistent with what was derived from the {\it
XMM-Newton} data (G08). First, we measure the radio
fluxes of the East lobe taking the same extraction region as for the
{\it Chandra}/{\it XMM-Newton} flux measurements, obtaining
$89.1$\,Jy, $31.7$\,Jy, $12.6$\,Jy, and $7.7$\,Jy, respectively for
$330$\,MHz, $1.4$\,GHz, $5$\,GHz, and $8.4$\,GHz radio maps. This
corresponds to a radio spectral index of $\alpha_{\rm R} \approx 0.75$.
Assuming a spherical volume with a radius $R = 36$\,kpc (corresponding
to $60''$; see \S\ref{sec:lobe}), the equipartition magnetic field
strengths for the East lobe is $B_{\rm eq} \simeq 10 $\,$\mu$G (e.g.,
Kataoka \& Stawarz 2005)\footnote{See equation (3) in Kataoka \&
Stawarz (2005), with minor changes made to mimic the IC/CMB model
presented in G08 and Croston et al (2005). In
particular, we assume here a broken power-law electron distribution
with initial electron energy index $s = 2$, and the minimum electron
Lorentz factor $\gamma_{\rm min} = 10$.}. By assuming that all the
lobe X-ray emission is due to inverse-Comptonization of the cosmic
microwave background photons, the measured flux densities imply a
magnetic field strength $B_{\rm ic} \simeq 6 $\,$\mu$G. Therefore, the
equipartition magnetic field intensity exceeds the value measured for
the East lobe by a factor $\sim$2. This result is indeed in a very
good agreement to what is observed in other FR~II sources (see Croston
et al. 2005, Kataoka \& Stawarz 2005, and references therein) and
consistent with what was measured by G08. Assuming further
equal energies stored in the non-thermal electrons and protons, the
total energy density of the 3C~353 lobes is $U_{\rm
lobe} \sim 3 \times10^{-11}$\,erg\,cm$^{-3}$. The total volume of the
lobes is $V_{\rm lobe} \sim 10^{70}$\,cm$^3$. This gives a total
energy $E_{\rm tot} \sim U_{\rm lobe} \times V_{\rm lobe} \sim 3
\times 10^{59}$\,erg, which is in fact a lower limit only. Finally, we
estimate the radio source lifetime to be $t_{\rm j} \sim
l_{\rm j} / v_{\rm adv} \sim 7 \times 10^6$\,yrs, which corresponds to the
observed jet length $l_{\rm j} \sim 2\arcmin$ and the assumed advance
velocity of the jet termination region $v_{\rm adv} \sim 0.03\,c$, as
appropriate for a typical FR~II radio galaxy (see, e.g., Machalski
et al. 2007). This constrains the jet kinetic power in the 3C~353
radio galaxy to be $L_{\rm j} \sim E_{\rm tot} / 2 \, t_{\rm j} \sim
10^{45}$\,erg\,s$^{-1}$.
 
The above order-of-magnitude calculations allow us to estimate the jet
magnetic field, $B$, by minimizing the total kinetic luminosity of the
3C~353 jets, $L_{\rm j} = L_{\rm p} + L_{\rm e} + L_{\rm B}$ (the
sum of the kinetic powers carried by the jet protons, electrons, and the
magnetic field, respectively), for a given jet synchrotron emission
(Ghisellini \& Celotti 2001). In particular, with
$L_{\rm B} = \pi R_{\rm j}^2 \, c \Gamma_{\rm j}^2 \, U_{\rm B}'$, where
$R_{\rm j} \approx 1.2$\,kpc is the jet radius (Swain et al. 1998),
$\Gamma_{\rm j}$ is the jet bulk Lorentz factor, and $U_{\rm B}' \equiv
B^2 / 8 \pi$ is the comoving energy density of the jet magnetic field,
one obtains
\begin{equation}
B = \left({ 4 \, L_{\rm j} \over R_{\rm j}^2 \, c \Gamma_{\rm
     j}^2}\right)^{1/2} \sim 100 \, \Gamma_{\rm j}^{-1} \, \mu\,{\rm G}
\, ,
\label{eq:b}
\end{equation}
\noindent
which is very close to the standard equipartition value (Swain
1996). From this, one may find the observer-frame ratio of the IC/CMB and
synchrotron luminosities as
\begin{equation}
{L_{\rm ic/cmb} \over L_{\rm syn}} \approx \left({\delta_{\rm j} \over \Gamma_{\rm j}}\right)^2 { U_{\rm cmb}' \over U_{\rm B}' } \sim 10^{-3} \, .
\label{eq:ratio}
\end{equation}
\noindent
Here, $U_{\rm cmb}' \approx 4 \times 10^{-13} \, \Gamma_{\rm
j}^2$\,erg\,cm$^{-3}$ is the jet comoving energy density of the CMB
radiation, $U_{\rm B}' \sim 4 \times 10^{-10} \,
\Gamma_{\rm j}^{-2}$\,erg\,cm$^{-3}$ (see equation \ref{eq:b}), and $\delta_{\rm j}
\equiv \left[ \Gamma_{\rm j} \, (1 - \beta_{\rm j} \cos \theta_{\rm j})\right]^{-1}$
is the jet Doppler factor.

Based on their models of the jet total intensity and polarization
profiles, Swain et al. (1998) constrained the jet inclination to the
line of sight in 3C~353 to be $60^{\circ} < \theta_{\rm j} <
90^{\circ}$; these large angles to the line of sight are strongly
supported by the observed two-sidedness of both the radio and X-ray
jets. Accordingly, we can approximately substitute $\delta_{\rm j}
\sim 1 / \Gamma_{\rm j}$ in equation \ref{eq:ratio}, obtaining the
observed luminosity ratio $ L_{\rm ic/cmb} / L_{\rm syn}$
independently of the jet kinematic factors. The resulting value, $\sim
10^{-3}$, is in strong disagreement with the observed X-ray-to-radio
luminosity ratio $L_{\rm X} / L_{\rm R} \gtrsim 1$ (see
Figure~\ref{fig:SED}). Note, that in the framework of the IC/CMB model
with large jet inclinations such that $\delta_{\rm j} \sim 1 /
\Gamma_{\rm j}$, the electrons producing inverse-Compton emission at
the observed keV photon energies have Lorentz factors $\gamma_{\rm X}
\sim (\nu_{\rm keV} / \nu_{\rm cmb} \, \delta_{\rm j}^2)^{1/2} \sim
10^3 \, \Gamma_{\rm j}$, i.e., almost exactly the same as the
electrons producing synchrotron photons at the observed GHz
frequencies, $\gamma_{\rm R} \sim (\nu_{\rm GHz} / 4.2 \times 10^6 \,
B \, \delta_{\rm j})^{1/2} \sim 10^3 \, \Gamma_{\rm j}$, so one can
write $ L_{\rm ic/cmb} / L_{\rm syn} \sim L_{\rm X} / L_{\rm R}$.
Thus, we conclude that the IC/CMB model cannot explain the observed
X-ray emission of the 3C~353 jets, unless very large departures from
energy equipartition, $B \ll 100 \, \Gamma_{\rm j}^{-1}$\,$\mu$G, are
invoked. Such large deviations from the minimum power condition are
not expected in the case of large-scale extragalactic jets (see in
this context Stawarz et al. 2005, 2006) and they would imply very high
total energy densities in the jet.

The only possibility left is therefore that the observed X-ray
emission of 3C~353 results from the synchrotron radiation of some
flat-spectrum high-energy electron population, most likely separate to
the one producing the observed radio emission. The required Lorentz
factors of the electrons emitting synchrotron photons with the
observed keV energies are $\gamma_{\rm X} \sim (\nu_{\rm keV} / 4.2
\times 10^6 \, B \, \delta_{\rm j})^{1/2} \sim 3 \times 10^7 \,
\Gamma_{\rm j}$ (assuming the scaling of the magnetic field as given
in equation 2 above and, again, $\delta_{\rm j} \sim 1 / \Gamma_{\rm
j}$). It has already been shown that stochastic acceleration processes
taking place in large-scale extragalactic outflows may easily account
for the production of electrons with these Lorentz factors (Stawarz \&
Ostrowski 2002, Uchiyama et al. 2006) and they are already invoked to
explain observations of the hotspots of FR~II sources and the jets of
FR~Is (\S\ref{sec:intro}). In the following section, by discussing in
detail the X-ray/radio morphology of the 3C~353 jets, we aim to
constrain several other aspects of the synchrotron scenario.

\subsection{Radio/X-ray Jet Morphology}
\label{sec:jet-physics}
Before discussing in more details the radio/X-ray morphology of the
3C~353 jets, it is useful to summarize the main results of the extensive radio
studies carried out on this source by Swain (1996) and Swain
et al. (1998). By means of detailed modeling of the intensity and
polarization properties of the jets in 3C~353, these authors concluded
that:
\begin{itemize}
\item most of the observed radio emission from the jet (as well as from
      the counterjet), is produced within the outer sheath of the
      outflow/jet boundary layer with a velocity shear; the contribution
      from the central jet spine is negligible;
\item the magnetic field within the jet boundary layer has no radial
      component, $B_r = 0$, but only random axial and toroidal
      components in equipartition, $\langle B_z^2 \rangle^{1/2} \approx
      \langle B_{\phi}^2 \rangle^{1/2}$, both ordered by a velocity
      shear; the axial component of the magnetic field within the jet
      spine is negligible;
\item the outer sheath and the jet spine have constant half-opening
      angle $\Theta \approx 0.5^{\circ}$ all along the outflow, such
      that the radius of the spine is roughly half of the total jet
      radius $R_{\rm j} \approx 2\arcsec \approx 1.2$\,kpc;
\item jet models with ordered large-scale magnetic field (helical
      structure, transverse flux ropes, etc.) cannot account for the
      observed flat-topped synchrotron intensity profiles across the
      jets and, at the same time, for the symmetric polarization
      profiles with measured degree of linear polarization $\sim
      10\%-20\%$.
\end{itemize}

\subsubsection{The Widths of the Radio and X-ray Jets}

As described in \S\ref{sec:jetwidth} of this paper, we have found that
the X-ray \emph{knots} in 3C~353 are possibly narrower than their
radio counterparts (which are also situated further away from the
active nucleus).
This agrees with what is observed in powerful quasar jets
detected by {\it Chandra}: the limits to the width of the
(transversely unresolved) X-ray knots in these systems are always slightly
smaller than the measured sizes in the radio band, and sometimes even
in the optical band (Jester et al. 2006, 2007). In the framework of
the IC/CMB model involving significant beaming, one would argue that
since the comoving energy density of the CMB photons is proportional
to the square of the bulk Lorentz factor of the emitting plasma,
$U'_{\rm cmb} \approx \Gamma_{\rm j}^2 \, U_{\rm cmb}$, the comoving
IC/CMB emissivity of the slower jet boundary layer should be much
smaller than the analogous emissivity of a fast jet spine, resulting
in the X-ray jet being narrower than its radio/optical counterpart
(for a fixed magnetic field energy density $U'_{\rm B}$ throughout
the outflow). However, as discussed in the previous section, the detection
of the X-ray counter-knot in 3C~353 and the established large
inclination of this source exclude the IC/CMB model: only the
synchrotron model is viable to explain these features. Thus, the
observations imply
that the presence of high-energy electrons producing the observed
synchrotron X-ray jet emission must be restricted to the
central spine of the jet rather than to the jet boundary layer,
\emph{at least in the upstream portions of the bright knots}, if our
findings regarding the widths of the X-ray jet features are correct.
We speculate that such a difference may be due to the different
configurations of the small-scale magnetic field within different parts
of the radially stratified outflow, as discussed by Swain et al.
(1998).

It should be borne in mind, however, that with the current data the
signal-to-noise ratio of the {\it Chandra} X-ray maps of 3C~353 is
rather poor, and therefore all conclusions regarding the width of the
X-ray jets in this object, or rather of the isolated jet knots, should
be taken with extreme caution. Also, we emphasize that the
multiwavelength structures and spectra of bright knots in other
well-studied jet sources (3C~273, Centaurus~A) do not necessarily
correspond directly to the multiwavelength morphology and spectral
properties of the interknot regions (Marshall et al. 2005b, Kataoka et
al. 2006, Hardcastle et al. 2007b), but that it is the interknot
regions (rather than the bright knots) that reflect closely the
global, possibly stratified morphology of the outflow. Finally, the
poorly known internal structure of the boundary layer in relativistic
jets may be much more complicated than what is typically assumed,
resulting in several non-intuitive radiative properties of the jet
plasma (see in this context Aloy \& Mimica 2008).

\subsubsection{Decrease of the $F_{\rm X}/F_{\rm R}$ Flux Ratio Along the Jets}
\label{sec:ratio-discussion}
In the case of the jets in 3C~353, we observe a
systematic decrease of the $F_{\rm X}/F_{\rm R}$ ratio along the
outflow (see \S\ref{sec:jet-spectra} and Figure \ref{fig:ratio}).
Similar behavior is observed in many (though not all) quasar jets
detected by {\it Chandra} (e.g., Sambruna et al. 2004, Marshall et al.
2005a, Hardcastle 2006, Siemiginowska et al. 2007). In the framework
of the IC/CMB model for the X-ray jet emission, it has been suggested
that this behavior results from a smooth deceleration of the outflows
on kpc-Mpc scales due to gradual mass entrainment (Georganopoulos \&
Kazanas 2004, Tavecchio et al. 2006). As pointed out by Hardcastle
(2006), this idea faces several difficulties in explaining the
multiwavelength morphology of powerful jets, and here we briefly
summarize the arguments given: (i) no trend for the gradual bulk
deceleration is observed in large-scale radio jets of FR~II sources,
but if only the jet spine is to be decelerated, the model should
explain why the jet boundaries do not respond to the involved dramatic
changes in the jet/spine kinematics; (ii) the amount of cold matter
required to be entrained, and thus to decelerate powerful quasar jets,
is in some cases too high when compared to the upper limits on the
masses of the surrounding thermal gas. As discussed previously, the
IC/CMB model is excluded in the case of 3C~353. On
the other hand, in the previous subsection we have argued that the
X-ray morphology implies that the synchrotron X-ray emission is
associated with the faster jet spine rather than with the slower jet
boundaries. Could the observed systematic decrease of the $F_{\rm
X}/F_{\rm R}$ flux ratio therefore nevertheless be connected with a
smooth jet/spine deceleration?

It is important to remark again in this context that the jets in
3C~353 are extremely well collimated all over their
lengths, with a constant half-opening angle $\Theta \approx
0.5^{\circ}$, and constant jet/counterjet brightness asymmetry ratio
$\mathcal{R}_{\rm R} \approx 2$. This is in agreement with what is
observed in other FR~IIs: no evidence for gradual jet deceleration is
provided by the jet radio morphology (Hardcastle 2006). Moreover, as
pointed out by Swain et al. (1998), the fact that the boundary layer
seems to be always restricted to the same parts of the outflow $> 0.5
\, R_{\rm j}$ (i.e., that the boundary shear layer does not
spread/expand toward the jet axis), coupled with the constancy of the
observed level of linear polarization along the jets and the absence
of radial component of the magnetic field, indicate that no
large-scale turbulent eddies associated with mass entrainment are
present in the 3C~353 jets. Since it is the jet boundary shear layer
which mediates interaction of the jet with the surrounding medium, the
absence of changes in the shear layer implies that there is no reason
for the jet spine to decelerate. Thus, we must conclude that
systematic mass entrainment on scales of tens of kpc is unlikely to be
of any importance, at least in 3C~353. The only remaining possibility
is thus that the particle acceleration conditions associated with
microphysical plasma parameters --- rather than with the global
hydrodynamical configuration of the outflow --- change within the
central parts of the jets. Because those regions of the jet dominate
the jet radiative output at X-ray frequencies but produce negligible
radio emission when compared to the jet boundaries, the ratio of the
observed X-ray and radio fluxes decreases systematically along the jet
in 3C~353 and in other analogous systems. We might for example
speculate that a gradually decaying chaotic component of the magnetic
field in the jet spine leads to a lower efficiency in the acceleration
(due to, e.g., magnetic reconnection processes) of the highest-energy
electrons.

\subsubsection{Positional Offsets Between the X-ray and Radio Knots}

The apparent positional offsets between the intensity maxima of the
X-ray and radio knots, as reported in this paper for 3C~353, are
observed in several other jet sources detected by {\it Chandra} (see,
e.g., Siemiginowska et al. 2007). As discussed by Hardcastle et al.
(2003), they cannot be explained by simple models involving particle
acceleration at a single extended shock front, which just exploit the
difference between the cooling timescales for the electrons emitting
synchrotron X-ray and radio photons (Bai \& Lee 2003). In a wider
context, the problem of such offsets relates directly to the question
on the nature of the jet knots: are they separate moving portions of
the jet matter, or rather stationary features produced by, e.g.,
reconfinement shocks? We note that formation of reconfinement shocks
in relativistic outflows has recently been widely discussed in the
context of the inner ($<1$\,kpc) jets of $\gamma$-ray emitting blazars
and radio galaxies (e.g., Jorstad et al. 2001, Cheung et al. 2007,
Levinson \& Bromberg 2008). Reconfinement shocks have been proposed to
explain the global morphology of kpc-scale FR~I jets (Laing \& Bridle
2002), and also --- which is the most interesting in the context of
our discussion --- the almost regularly spaced knots in FR~II jets on
scales of tens to hundreds of kpc (Komissarov 1994, Komissarov \&
Falle 1998). On the other hand, the frequency independence of the longitudinal
profiles of the knots in the quasar jets observed by {\it Chandra},
among other observed jet properties, has led several authors to
conclude that knots are instead separate moving portions of the jet
with excess kinetic power, produced by intermittent or modulated
activity of the central engine (Bridle et al. 1986, 1989, Clarke et
al. 1992, Stawarz et al. 2004).

The results of the analysis of the radio data for the
3C~353 jets presented by Swain et al. (1998), in particular the reported
constant opening angle of \emph{both} the jet spine and the jet boundary
layer, is hard to reconcile with the expected `diamond structure' of the
successive reconfinement shocks discussed by Komissarov \& Falle
(1998). In addition, the reconfinement shock scenario does not predict in a
natural way the frequency-dependent position of the intensity maxima of the
radiating plasma. Clearly, this problem cannot be addressed fully in our
qualitative discussion, since detailed numerical simulations are needed
to understand the complex structure of the reconfinement shocks in
large-scale mildly-relativistic outflows. However, we believe that a
more likely explanation for the observed X-ray/radio positional offsets
can be given in the framework of the model in which knots are moving
portions of the jet matter. For example, one may propose that portions
of the jet produced by the central engine during a more active period,
and hence characterized by excess kinetic power and/or
higher bulk velocity, propagate within the `stationary' (slower) outflow
corresponding to the quasi-quiescent state of the active nucleus. Since
the bulk velocities of the jet plasma are expected to be supersonic
(although not necessarily ultrarelativistic), a double-shock structure is
expected to form, due to collisions of the faster portions of the jet
matter with the slower outflow.

Let us consider a model in which the forward shock
propagating in the slower outflow (and thus characterized by a highly
oblique geometry at the jet boundaries) is responsible for the brightening
of the outer portions of the jet at radio frequencies (due to
compression of the jet magnetic field, driving the turbulence, etc.),
while the reverse shock propagating within the faster portion of the jet
is associated with the intensity maximum of the synchrotron X-ray
emission. In this case we may make some simple illustrative
calculations of the expected magnitude of the offsets. Let $\Gamma_1$ be the bulk Lorentz factor of the slower
(upstream) portion of the outflow, and $\Gamma_2$ be the bulk Lorentz
factor of the faster (but otherwise identical) portion of the jet. After
the collision, a symmetric double-shock structure forms within the
outflow, for which the contact discontinuity propagates with the
velocity $\beta_{\rm j} = (1 - \Gamma_{\rm sh}^{-2})^{1/2}$ in the
observer's frame. This is, of course, the bulk velocity of the radiating
(downstream) jet plasma. Let us also assume that the slow portion of the
outflow is at most mildly relativistic (see the discussion in the
previous sections), i.e., that $\Gamma_1 \sim 1$. With this assumption, the
velocity of the shock in the observer frame is $\beta_{\rm sh} = (1 -
\Gamma_{\rm sh}^{-2})^{1/2}$, where
\begin{equation}
\Gamma_{\rm sh}^2 = {\left(\Gamma_{\rm j} + 1\right) \, \left[
		      \hat{\gamma} \, \left(\Gamma_{\rm j}-1\right) +
		      1\right]^2 \over \hat{\gamma} \,
\left(2-\hat{\gamma}\right)\,\left(\Gamma_{\rm j}-1\right) + 2} \, ,
\end{equation}
\noindent
and $\hat{\gamma}$ is the ratio of the specific heats of the jet plasma
(see, e.g., Stawarz et al. 2004, Appendix E). The extent of the shocked
region in the downstream plasma rest frame (denoted by primes) is
$\Delta l' = 2 c \, \beta_{\rm sh}' \, \Delta t'$, where $\beta_{\rm
sh}' = (\beta_{\rm sh}-\beta_{\rm j})/(1-\beta_{\rm sh} \, \beta_{\rm
j})$, and $\Delta t'$ is the comoving time since collision. Because for
the moving source $\Delta l = \delta_{\rm j} \, \Delta l'$ and $\Delta t
= \Gamma_{\rm j} \, \Delta t'$, we obtain the expected separation of
the radio and X-ray intensity maxima (i.e., the separation between the
forward and reverse shock) 
\begin{equation}
\Delta l = 2 c \, \beta_{\rm sh}' \, \Gamma_{\rm j}^{-2} \, \Delta t \, ,
\end{equation}
\noindent
where we take $\delta_{\rm j} \equiv 1 / \Gamma_{\rm j} \, (1-\beta_{\rm
j} \, \cos \theta_{\rm j}) \sim 1 / \Gamma_{\rm j}$ as appropriate for
a jet viewed at large inclinations $\theta_{\rm j} \lesssim
90^{\circ}$. Note that in the framework of the proposed toy model, one
should expect the spatial offset $\Delta l$ to increase with $\Delta t$,
that is, with the distance from the nucleus. This is in agreement with
what is observed in 3C~353. Note also that the timescale required to
explain the kpc-scale positional differences between the intensity
maxima of the knots, $\Delta t \approx 3 \times 10^4 \, (\Delta l / {\rm
kpc})$\,yrs, is in a very good agreement with what has been proposed for the
intermittent activity of the central engine in powerful FR~II/quasar
sources (Reynolds \& Begelman 1997, Siemiginowska \& Elvis 1997, 
Stawarz et al. 2004). In the above estimate we have
assumed a very moderate bulk Lorentz factor for the jet $\Gamma_{\rm j} =
2$ (corresponding to a bulk velocity of the radiating plasma $\beta_{\rm j} = 0.866$), and
the ratio of the specific heats $\hat{\gamma} = 4/3$. For these
parameters we find $\Gamma_{\rm sh} = 2.38$, $\beta_{\rm sh} =
0.907$, and $\beta_{\rm sh}' = 0.193$.

Several features of the observations are not explained by this
scenario; for example, (i) there is no obvious reason why the forward
and reverse fronts of the double shock structure should work
substantially differently with respect to the acceleration of
ultrarelativistic electrons; and (ii) it is not clear why similar
positional offsets are observed in the jet knots and in the jet
termination regions (hotspots), since the shock structures produced in
the latter regions are expected to be completely different. These may
suggest that the toy model considered above is over-simplified, and
that some other physical picture has to be considered. Another
possibility --- still in the framework of intermittent jet activity
--- is that the central engine ejects \emph{heavy} knots in $\sim
10^4$\,yr-long flare-like activity periods. Due to the high
(relativistic) mass load of the knot, the bulk velocity of one of
these heavier portions of magnetized plasma is expected to be smaller
than the velocity of the `weak jet' formed in an epoch of quiescence.
Interactions between the heavy/slower and light/faster portions of the
jet would lead to formation of a single (reverse) shock at the
upstream edge of a heavy knot (the forward shock, propagating within
the heavy portion of the jet matter, would be expected to die away
soon after its formation, as long as the density ratio between the two
phases of the ejecta was high). Thus, the heavy portion of the jet
outflow would behave like the reflecting walls used in
`particle-in-cell' simulations and would form a shock front that would
propagate upstream. As a result, the positions of radio knots would
coincide with the positions of heavy portions of the jet plasma, while
the X-ray intensity maxima due to freshly accelerated electrons would
be expected to coincide with the downstream region of the reverse
shock. This model solves at least problem (i) noted above; however,
discussions based on more sophisticated numerical simulations are
clearly essential to confirm this or other similar speculations.

\section{Conclusions}
\label{sec:conclusion}

We have presented a detailed analysis of the data for the 
powerful FR~II radio galaxy 3C~353 obtained with the {\it Chandra} X-ray
Observatory. In a deep, $90$\,ks {\it Chandra} observation, we have 
detected the X-ray emission from most of the radio structures in this source, 
including the nucleus, isolated knots in the jet and the counterjet,
the terminal jet regions (hotspots), and one radio lobe. Our major findings 
are as follows:
\renewcommand\theenumi{(\roman{enumi})}
\begin{enumerate}
\item Non-thermal X-ray emission associated with a knot in the 
      counterjet of a powerful FR~II source has been detected for the first time.
      This detection, in agreement with the established large inclination
      of 3C~353, strongly disagrees with the 
      inverse-Compton model proposed in the literature, and points to 
      a synchrotron origin for the X-ray jet photons. 
\item We find that the width of the X-ray knots is narrower
      ($\gtrsim 4\sigma$ effect) than that measured at radio
      wavelengths, suggesting that the production of the X-ray emission is  
      associated with the central jet spine rather than with the jet 
      boundaries. This conclusion, however, must be confirmed with
      X-ray data with a much better signal-to-noise ratio.
\item The radio-to-X-ray flux ratio decreases systematically downstream
      along the outflows, as is often observed in other extragalactic
      sources. We argue that this is due to the fact that the particle 
       acceleration conditions associated with microphysical plasma
      parameters changes within the central part of the jet,
      rather than being due to a gradual decrease of the jet bulk
      velocity resulting from a smooth jet deceleration.
\item Substantial (kpc-scale) offsets between positions of the X-ray and 
      radio intensity maxima within each knot are found, with the
      magnitude of the offsets increasing away from the nucleus. We have 
      speculated that these offsets can be explained if radio knots are 
      moving portions of jet material which are produced by the central 
      engine during an epoch of enhanced activity. Depending on the
      velocity and density ratio between these knots and the outflow
      produced at times of quiescence, a complex double-shock
      structure may form as a result of the interaction between the two 
      phases of the ejecta, with the reverse shock (propagating within 
      the faster portion of the jet) being associated with the peak of the
      X-ray emission.
\end{enumerate}
Although we cannot provide definitive solutions or interpretations for
each problem, we argue that the 3C~353 data strongly suggest that the
synchrotron X-ray emission of extragalactic large-scale jets is not 
simply shaped by the global hydrodynamical configuration of the
outflows, but is also very sensitive to the microphysical parameters
of the jet plasma.

\acknowledgments

J.K.\ acknowledges support by JSPS KAKENHI (19204017/14GS0211).
{\L}.S. and M.O. acknowledge support by the MEiN grant 1-P03D-003-29.
M.J.H. acknowledges support from the Royal Society and J.L.G. thanks
the UK Science and Technology Facilities Council for a studentship.
M.O. thanks Malgosia Mochol for her assistance in the evaluation of
the {\it Chandra} data. This research is funded in part by NASA 
contract NAS8-39073.  Partial support for this work was provided by 
the National Aeronautics and Space Administration through Chandra 
Awards Number GO5-6113X and GO7-8103X-R issued by the Chandra 
X-Ray Observatory Center, which is operated by the Smithsonian 
Astrophysical Observatory for and on behalf of NASA under contract 
NAS8-39073. This research has made use of data obtained by Chandra 
X-ray Observatory and software provided by the Chandra X-ray Center (CXC).
This research has also made use of SAOImage DS9,
developed by Smithsonian Astrophysical Observatory. The National Radio
Astronomy Observatory is a facility of the National Science Foundation
operated under cooperative agreement by Associated Universities, Inc.

\end{document}